\documentclass[a4paper,11pt]{article}

\usepackage{amssymb,lscape}
\usepackage{multirow}
\usepackage{amscd,bbold}
\usepackage{amsmath}
\usepackage{cite}
\usepackage{epstopdf}
\usepackage{xypic}
\usepackage[matrix,arrow,curve]{xy}

\numberwithin{equation}{section}       

\allowdisplaybreaks                    

\newcommand{\beq}{\begin{equation}}
\newcommand{\eeq}{\end{equation}}
\newcommand{\ba}{\begin{align}}
\newcommand{\ea}{\end{align}}
\newcommand{\baed}{\begin{aligned}}
\newcommand{\eaed}{\end{aligned}}
\newcommand{\bea}{\begin{eqnarray}}
\newcommand{\eea}{\end{eqnarray}}
\newcommand{\nn}{\nonumber}
\newcommand{\dd}{\mathrm{d}}
\newcommand{\ee}{\mathrm{e}}
\newcommand{\ii}{\mathrm{i}}

\newcommand{\del}{\partial}

\newcommand{\bbR}{\mathbb{R}}
\newcommand{\bbC}{\mathbb{C}}

\DeclareMathOperator{\SU}{\mathit{SU}}
\DeclareMathOperator{\SO}{\mathit{SO}}

\DeclareMathOperator{\GL}{\mathit{GL}}

\DeclareMathOperator{\Spin}{\mathit{Spin}}
\DeclareMathOperator{\Pin}{\mathit{Pin}}

\newcommand{\aaa}{{\cal A}}
\newcommand{\bbb}{{\cal B}}
\newcommand{\ccc}{{\cal C}}
\newcommand{\ddd}{{\cal D}}

\newcommand{\id}{\mathbb{1}}

\DeclareMathOperator{\re}{Re}

\newcommand{\stt}{\SU(3)\times\SU(3)}
\newcommand{\kpara}{k_{\parallel}}
\newcommand{\kperp}{k_{\perp}}

\newcommand{\te}{\tilde{e}}

\newcommand{\cE}{\mathcal{E}}
\newcommand{\tcE}{\Tilde{\cE}}
\newcommand{\ttcE}{\Tilde{\Tilde{\cE}}}
\newcommand{\Ggeom}{G_{\mathrm{geom}}}
\newcommand{\Lgen}{\mathbb{L}}

\newcommand{\DD}{\slash{\mspace{-13mu}D}}

\newcommand{\gO}{\mathbb{\Omega}}

\begin{document}


\begin{titlepage}

\begin{center}

\rightline{\small SPhT-T08/118}
\rightline{\small Imperial/TP/08/DW/02}
\vskip 2cm

\begin{LARGE}
   \textbf{T-duality, Generalized Geometry and Non-Geometric \\*[5pt]
     Backgrounds}
\end{LARGE}

\vskip 1.2cm

\textbf{Mariana Gra{\~n}a$^{a}$, Ruben Minasian$^{a}$, Michela
  Petrini$^{b}$ and Daniel Waldram$^{c}$}

\vskip 0.8cm
{}$^{a}$\textit{Institut de Physique Th\'eorique,                   
CEA/Saclay \\
91191 Gif-sur-Yvette Cedex, France}  \\

\vskip 0.4cm
{}$^b$ \textit{LPTHE, Universit\'es Paris VI et VII, Jussieu \\
75252 Paris, France} \\

\vskip 0.4cm
{}$^{c}$\textit{Department of Physics and Institute for Mathematical 
  Sciences\\
  Imperial College London, London, SW7 2BZ, U.K.}\\

\end{center}

\vskip 1cm

\begin{center} {\bf Abstract } \end{center}
\begin{quote}
We discuss the action of $O(d,d)$, and in particular T-duality, in the
context of generalized geometry, focusing on the description of so-called
non-geometric backgrounds.  We derive local expressions for the pure
spinors descibing the generalized geometry dual to an $\SU(3)$
structure background, and show that the equations for
$N=1$ vacua are invariant under T-duality. We also propose a
local generalized geometrical definition of the charges $f$, $H$, $Q$
and $R$ appearing in effective four-dimensional theories, using the
Courant bracket. We then address certain global aspects, in particular
whether the local non-geometric charges can be gauged away in, for instance,
backgrounds admitting a torus action, as well as the structure of
generalized parallelizable backgrounds.    
\end{quote}

\vfill
\today

\end{titlepage}


\tableofcontents


\section{Introduction}

Recent developments in flux compactifications brought T-duality to the
center stage
\cite{Hull,HullRE,Taylor,nongeomflux,Taylor2,nongeovacua,GLW2,PS1,PMT,dallprezas}. Given
a background with isometries, T-duality is a very effective tool for 
generating new backgrounds. Due to the  mixing of the metric and NS
two-form $B$, it can relate string backgrounds with drastically
different properties.  

A well known example is the action of a single T-duality along an
isometry direction of a manifold with an $H$ flux. At the local
level, T-duality exchanges the off-diagonal components of the metric
with those of the $B$ field. Since the Killing vector generating the
isometry must be globally defined, the manifold can be thought of as a
circle fibration over a base $M$. Topologically the fibration can be
characterized by the first Chern number. Another topological number is
given by the integral over a two-cycle on the base of the two-form
obtained by contracting the $H$-flux with the Killing
vector. T-duality exchanges these numbers, leading to a change in
topology \cite{Mathai}. Indeed, the dual manifold is again a circle
fibration over the same base, but  has in  general a different first
Chern number.

For backgrounds with $d$ isometries, since the T-duality group $O(d,d)$ is much larger than the group of diffeomorphisms, it is not surprising that 
the action of T-duality can be more interesting and complicated \cite{Mathai2}. 

At the level of string sigma model there is a well-defined procedure
of performing the duality transformations  (see e.g. \cite{gpr}). Given a background with
isometries, one gauges them and adds Lagrange multipliers. Integrating
out the original directions of isometries, while leading to the
same two-dimensional quantum field theory, yields a different target
space. There are however global obstructions in performing the above
procedure \cite{Hull:2006qs, BHM}. In order to perform T-duality 
(in three or more directions) the component of the 
$H$ flux fully lying in the directions to be dualized 
has to vanish, in other words, the $B$-field must respect the isometries.
Also, the component of $H$ with two legs along the duality directions must  
be trivial in cohomology, i.e. the corresponding component of the $B$-field is globally defined.
A priori, when such global obstructions are present, T-duality is not possible. 

However, it has been proposed in \cite{Hull:2006qs}, that in some of the obstructed
cases, T-duality can lead to consistent string backgrounds.  
While it is still possible to give local expressions for the metric and 
$B$-field,  globally the resulting background will not have a conventional
description as a good internal manifold, thus the terming ``non-geometric compactification''.

The interest in such ``non-geometric'' backgrounds is also motived by the analysis
of four-dimensional effective theories. From the 
point of view of gauged supergravities in four
dimensions, potentials governed by a large duality group seem to admit
various minima, many of which cannot correspond to conventional string
compactifications. It is interesting to identify those that
can be lifted to full string solutions, and the backgrounds obtained via  
obstructed T-duality transformations are natural candidates. There is much recent work 
supporting this possibility \cite{Taylor2,nongeovacua}. When trying to
lift solutions of four-dimensional gauged supergravities to ten
dimensions, the origin of the gauged symmetries as well as that of the
structure constants in their algebra need to be given a string theory
interpretation. For compactifications on $d$-dimensional homogeneous
parallelizable manifolds (loosely called twisted tori), there are $2d$
symmetries corresponding to translations and  to gauge transformations
on $B$. The twisting of the frame bundle and the $H$-flux appear as
structure constants of the ``Kaloper-Myers algebra'' ~\cite{KM}. This
algebra is however not covariant under the duality group $O(d,d)$,
since such an invariance would require twice as many charges. It has
been argued that the missing half corresponds to ``non-geometric
fluxes'', encoding, for example, monodromies in the T-duality group,
mixing metric and $B$-field \cite{HullRE,Taylor}.

The T-duality group $O(d,d)$ is also the structure group of the
generalized tangent bundle, which combines the tangent and cotangent
bundle of a $d$-dimensional manifold in Generalized Geometry
\cite{hitchin1,Gualtieri}. The generalized metric on the generalized
bundle encodes the information about the metric and $B$-field of the
manifold, which are exchanged by T-duality. Additionally, 
non-trivial patchings of the $B$-field are naturally incorporated in
generalized geometry by defining an extension of the tangent bundle by
the cotangent one. In this extension, the patching between two overlapping
regions uses, besides the usual diffeomorphisms, an abelian subgroup
of $O(d,d)$ involving the $B$-field. This suggests the use of
generalized geometry  to describe the action of T-duality
\cite{Gualtieri,TdGCG}.  

Generalized (complex) geometry is very well suited to describe ${\cal
  N} =1$ supersymmetric compactification with non trivial fluxes. When
there are nowhere vanishing spinors on the manifold, one can construct
bispinors  by  
tensoring two $O(d)$ spinors with the same and with opposite chiralities. These $O(d,d)$ spinors
also carry the information about the metric and B-field on the manifold.  Each $O(d,d)$ spinor 
corresponds to an algebraic structure, namely, for six-dimensional manifolds, an 
$SU(3,3) \subset O(6,6)$. The pair of $SU(3,3)$ structures defines an $\stt$ structure on the generalized tangent bundle. In the context of flux compactifications this can 
can be understood as two independent $SU(3)$ structures, one  on the left and one on the 
right moving sector. 
The corresponding globally defined $O(d)$ spinors are the internal supersymmetry parameters.
The conditions for ${\cal N}=1$ vacua 
reduce to a couple of first order differential equations 
for the pure spinors, implying the closure of one pure spinor and relating  the failure of 
integrability of the other to the RR fluxes \cite{GMPT2}.  
A manifold admitting a closed pure spinor is  a Generalized Calabi-Yau.

Even though generalized geometry conventionally describes ordinary
geometrical back--grounds, we will argue that it is still a
suitable language to describe some aspects of non-geometric
backgrounds. Specifically one is interested in the cases where the
compactification remains locally a manifold, such as arises from
obstructed T-duality of conventional geometries (some progress in this
direction has been done in \cite{PS1,PS2}).   

In this paper we discuss the action of $O(d,d)$ as well as the emergence of 
the extended Kaloper-Myers algebra, in the context of generalized geometry.
We will first do this at the local level, and then discuss the global properties. It is at this
point that the difference between geometry and non-geometry appears.
At the local level, we find the T-duality action on the objects defining the generalized metric,
namely the generalized vielbeins in the generic case, and the pure spinors in the case of 
reduced structure. 
We will mostly consider situations  where the $O(d,d)$ transformations are
along isometries of the background. We will assume that 
the $B$-field respects the isometries and yet is  not globally well-defined (in particular the 
components to be fully T-dualized). In these cases we argue 
that  T-duality yields perfectly good local expressions (mixing the metric and the $B$-field). 

We also show that the charges in the extended Kaloper-Myers algebra have a 
simple interpretation as 
elements of a generalized spin-connection, or, equivalently, as structure constants (or rather 
functions, in the generic case) of the Courant algebra of generalized vielbeins. 
As such, the distention between geometric and non-geometric charges depends on the frame, 
and therefore loses physical content. However, when turning to global properties, we show 
that the transformation taking from one frame to another can be ill-defined if there are 
non-contractible loops. In this case the non-geometric charges cannot be globally 
gauged away. The distinction between geometric and non-geometric situations 
is nicely rephrased in terms of right and left mover sectors
of string theory.  
Generalized geometry suggests the use of a different
set of vielbeins for the left and right movers, which transform nicely under $O(d,d)$.\footnote{Two sets of vielbeins on the target space were introduced in the context of T-duality in \cite{hassan}. They also appear naturally in the doubled formalism \cite{Hull}.}   
For geometric backgrounds it is always possible, after T-duality, to perform a well
defined local $O(d) \times O(d)$ transformation to set the vielbeins for the right and left 
moving sectors to be the same everywhere. On the contrary, when such a
transformation is only possible locally, the background is
non-geometric. A similar situation arises in the doubled formalism,
where for geometrical backgrounds one is again able to use $O(d) \times O(d)$
transformations to write the doubled vielbeins in a particular
triangular form\cite{dallprezas}.  

The two different sets of vielbeins on right and left movers have nice
transformation properties under $O(d,d)$ . Moreover they provide a way to determine
the $O(d,d)$ transformation of an $\stt$ structure. In principle, 
given the two new $\SU(3)$ structures, one should be able to build the corresponding T-dual
spinors. However, since the pure spinors  are mixtures of left and 
right moving sectors, determining them explicitly can be quite challenging. 
Here we use a different approach and we study directly the action of $O(d,d)$ on the spinors.
In particular we derive local expressions for the pure spinors dual, or mirror, to those 
corresponding to a single $\SU(3)$ structure (i.e. where the original structures on the left and on the right sector coincide). These correspond 
generically to $\stt$ structures. 

As already mentioned, generalized geometry allows for an elegant classification of
Type II flux background with ${\cal N}=1$ supersymmetry.
Given that supersymmetry equations are local, they could still be considered in the case
of non-geometric backgrounds, when these admit a local description. 
In other words, one might wonder whether considering
T-duality along isometries that commute with the supersymmetry generators, one might still
have good local solutions, also when T-duality is obstructed.
We show that the ${\cal N}=1$ supersymmetry equations on pure spinors are
invariant under T-duality. 

A very simple example of a generalized geometrical background is
one where there is a globally defined set of generalized vierbein, the
analogue of a conventional parallelizable background. The Courant
bracket on the preferred frame then provides a natural global
definition of the generalized charges. Many of the simplest
non-geometrical examples are of this ``generalized parallelizable''
type. We discuss some necessary conditions on the local geometry in
this case and in particular show that the $R$ charge always vanishes. 

The paper is organized as follows. In Section \ref{sec:gen-geom} we
review the necessary ingredients of generalized geometry, and find the
$O(d,d)$ transformations of the vielbeins. In Section
\ref{sec:Tduality} we discuss how T-duality acts on the generalized
structures, and find explicit expressions for the duals of an $SU(3)$
structure. We also show that the equations for ${\cal N}=1$ vacua are
invariant under T-duality. 
In Section \ref{sec:charges} we introduce the generalized spin connection
and we discuss how the charges of the extended Kaloper-Myers algebra
arise locally from the Courant bracket. Finally in Section
\ref{sec:global} we discuss global issues and non-geometricity, as
well as the structure of generalized parallelizable backgrounds.


\section{Generalized geometry}
\label{sec:gen-geom}


This section starts with a review of generalized geometry, the
generalized $O(d,d)$ spinors and the generalized metric ${\cal H}$, which
encodes the ordinary metric $g$ and the $B$-field. The generalized
metric defines an $O(d)\times O(d)$ structure, and we also introduce a
natural set of generalized vielbeins for ${\cal H}$. This latter has been
previously analyzed by Hassan~\cite{hassan}. One new element here is the
discussion of how the dilaton naturally enters the definition of
$O(d,d)$ spinors.  

We then turn, in the context of six-dimensional manifolds, to the
various definitions of $\stt$ structures relevant to supersymmetric
backgrounds. 

Of particular interest is how the $O(d,d)$ group acts on the
generalized vielbeins and hence on the ordinary vielbein and
$B$-field (also discussed in~\cite{hassan}). In addition we consider
the action of $O(6,6)$ on an $\stt$ structure. These will be useful in 
the following parts of the paper were we specialize to T-duality
transformations in a number of specific cases. Here we consider only
the action of $O(d,d)$ at a point in the manifold. 


\subsection{Generalized tangent bundle}
\label{sec:gengeom}

The basic idea of generalized geometry~\cite{hitchin1,Gualtieri} is to
combine vectors and one-forms into a single object. Formally, on a
$d$-dimensional manifold $M$ one introduces the generalized tangent
bundle $E$ which is a particular extension of $T$ by $T^*$
\begin{equation}
\label{eq:Edef}
   0 \longrightarrow T^*M \longrightarrow E 
      \stackrel{\pi}{\longrightarrow} TM \longrightarrow 0 . 
\end{equation}
Sections of $E$ are called generalized vectors. Locally they can be
written as $X=x+\xi$ where $x\in TM$ and $\xi\in T^*M$. In going from
one coordinate patch $U_\alpha$ to another $U_\beta$, we have to first
make the usual patching of vectors and one-forms, and then give a
further patching describing how $T^*M$ is fibered over $TM$ in
$E$. This gives
\begin{equation}
\label{eq:S-patch}
   x_{(\alpha)} + \xi_{(\alpha)}
      = a_{(\alpha\beta)} x_{(\beta)} 
         + \left[ a_{(\alpha\beta)}^{-T}\xi_{(\beta)}
            - i_{a_{(\alpha\beta)}x_{(\beta)}}\omega_{(\alpha\beta)}
         \right] \, ,
\end{equation}
where $a_{(\alpha\beta)}\in\GL(d,\bbR)$, $\omega_{(\alpha\beta)}$
is a two-form and $a^{-T}=(a^{-1})^T$. Using a two-component
notation to distinguish the vector and form parts of $X$ we can write
\begin{equation}
\label{eq:patch}
   X_{(\alpha)}
      = \begin{pmatrix} x_{(\alpha)} \\ \xi_{(\alpha)} \end{pmatrix} 
      = \begin{pmatrix} 
            \id & 0 \\ \omega_{(\alpha\beta)} & \id
         \end{pmatrix}
         \begin{pmatrix}
            a_{(\alpha\beta)} & 0 \\ 0 & a^{-T}_{(\alpha\beta)}
         \end{pmatrix}
         \begin{pmatrix} x_{(\beta)} \\ \xi_{(\beta)} \end{pmatrix} 
      = p_{(\alpha\beta)} X_{(\beta)} \, . 
\end{equation}
In fact one makes the further restriction that
$\omega_{(\alpha\beta)}=-\dd\Lambda_{(\alpha\beta)}$, where
$\Lambda_{(\alpha\beta)}$ are required to satisfy  
\begin{equation} 
\label{cocycle}
   \Lambda_{(\alpha\beta)} + \Lambda_{(\beta\gamma)} 
         + \Lambda_{(\gamma\alpha)} 
      = g_{(\alpha\beta\gamma)}\dd g_{(\alpha\beta\gamma)} 
\end{equation}
on $U_\alpha\cap U_\beta\cap U_\gamma$ and
$g_{\alpha\beta\gamma}:=\ee^{\ii\alpha}$ is a $U(1)$
element. This is analogous to the patching of a $U(1)$ bundle,
except that the transition ``functions'' are one-forms,
$\Lambda_{(\alpha\beta)}$. Formally it is called the ``connective
structure'' of a gerbe. The point is that it is the geometrical
structure one needs to introduce $B$, the two-form analogue of an
ordinary one-form $U(1)$ connection, with a correspondingly quantized
field strength $H$. 

Given the split into vectors and forms, there is a natural
$O(d,d)$-invariant metric $\eta$ on $E$, given, on each patch, by 
\begin{equation}
\label{eq:eta}
   \eta(X,X) = i_x\xi ,
\end{equation}
or, in two-component notation, $\eta(X,X)=X^T\eta X$ with  
\begin{equation}
   \eta = \frac{1}{2}\begin{pmatrix} 0 & \id \\ \id & 0 \end{pmatrix} .
\end{equation}

The metric is invariant under $O(d,d)$ transformations acting on the
fibres of $E$. A general element $O\in O(d,d)$ can be
written in terms of  $d \times d$ matrices $a$, $b$, $c$, and $d$ as 
\begin{equation}
\label{eq:oddel}
   O = \begin{pmatrix} a & b \\ c & d \end{pmatrix} \ ,
\end{equation}
under which a generic element $X \in E$ transforms by    
\begin{equation}
\label{eq:Odd}
\begin{aligned}
   X = \begin{pmatrix} x \\ \xi \end{pmatrix}
      &\mapsto OX = \begin{pmatrix} a & b \\ c & d \end{pmatrix}
         \begin{pmatrix} x \\ \xi \end{pmatrix} .
\end{aligned}
\end{equation}
The requirement that $\eta(OX,OX)=\eta(X,X)$ implies $a^T c +c^T a=0$,
$b^T d + d^T b=0$ and $a^T d+c^T b=\id$. Note that the $\GL(d)$ action on
the fibres of $TM$ and $T^*M$ embeds as a subgroup of $O(d,d)$.
Concretely it maps  
\begin{equation}
\label{eq:diffeo}
   X \mapsto X' = 
      \begin{pmatrix} a & 0 \\ 0 & a^{-T} \end{pmatrix}
      \begin{pmatrix} x \\ \xi \end{pmatrix} ,
\end{equation}
where $a\in\GL(d)$. Similarly the first factor in~\eqref{eq:patch} is
also an (Abelian) subgroup $G_B\subset O(d,d)$. Given a two-form
$\omega$, we write  
\begin{equation}
\label{eq:Btransform}
   \ee^\omega = 
      \begin{pmatrix} \id & 0 \\ \omega & \id \end{pmatrix} 
   \qquad \text{such that} \qquad
   X = x + \xi \mapsto X' = x + (\xi-i_x\omega) .
\end{equation}
This is usually referred to as a $B$-transform. Given a bivector
$\beta$ one can similarly define another Abelian subgroup of
$\beta$-transforms 
\begin{equation}
\label{eq:betatransform}
   \ee^\beta = 
      \begin{pmatrix} \id & \beta \\ 0 & \id \end{pmatrix} 
   \qquad \text{such that} \qquad
   X = x + \xi \mapsto X' = (x + \beta\cdot\xi)+  \xi  .
\end{equation}

The patching~\eqref{eq:patch} of $E$ was by elements of $\GL(d)$ and
$G_B$. Together these form a subgroup which is a semi-direct product
$\Ggeom=G_B\rtimes\GL(d)$. A general element of $\Ggeom$ can be
written as  
\begin{equation}
\label{eq:ggeom}
   p = \ee^\omega\begin{pmatrix} a & 0 \\ 0 & a^{-T} \end{pmatrix}
       =  \begin{pmatrix} a & 0 \\ \omega a & a^{-T} \end{pmatrix} .
\end{equation}
This patching means that the structure group of the generalized
tangent space $E$ actually reduces from $O(d,d)$ to $\Ggeom$. The embedding
of $\Ggeom\subset O(d,d)$ is fixed by the projection $\pi:E\to TM$. It is
the subgroup which leaves the image of the related embedding $T^*M\to
E$ invariant.

There is also a natural bracket on generalized vectors known as the
Courant bracket, which encodes the differentiable structure of $E$ and
will play an important role in what follows. It is defined as 
\begin{equation}
\label{eq:Courant}
   [ x + \xi, y + \eta ] 
     = [x,y]_{\rm Lie} + \mathcal{L}_x\eta - \mathcal{L}_y\xi
        - \tfrac{1}{2} \dd\left(i_x\eta - i_y\xi\right) ,
\end{equation}
where $[x,y]_{\rm Lie}$ is the usual Lie bracket between vectors and
$\mathcal{L}_x$ is the Lie derivative. The Courant bracket is invariant under
the action of elements of $\Ggeom$, (\ref{eq:ggeom}), where
the $\GL(d)$ transformations $a$ are generated by diffeomorphisms and
the $B$-shifts $\omega$ are closed, $\dd\omega=0$.


\subsection{Generalized metrics, generalized vielbeins and $O(d)\times
  O(d)$ structures} 
\label{sec:vielbeins}

In the generalized geometry picture the metric $g$ and the $B$-field
combine into a single object which defines an $O(d)\times O(d)$
structure on $E$. To define an $O(d)\times O(d)$ structure we need the
bundle $E$ to split into two orthogonal $d$-dimensional sub-bundles
$E=C_+\oplus C_-$ such that the metric $\eta$ decomposes into a
positive-definite metric on $C_+$ and a negative-definite metric on
$C_-$. The subgroup of $O(d,d)$ which preserves each metric separately
is then $O(d)\times O(d)$. Since any element of $E$ which is a pure
vector or a pure one-form is null with respect to $\eta$, such
elements cannot lie in 
$C_+$ or $C_-$. Hence we can write a generic element $X_+\in C_+$ as
$x+Mx$, where $x\in TM$ and, in components, the form part is given by
$M_{mn}x^n$ for some general matrix $M$. (This actually describes an
isomorphism between $TM$ and $C_+$.) If we write
$M_{mn}=B_{mn}+g_{mn}$, where $g$ is symmetric and $B$ antisymmetric,
we see that the patching condition~\eqref{eq:patch} implies that
\begin{equation} \label{Bpatch}
   g_{(\alpha)} = g_{(\beta)} , \qquad
   B_{(\alpha)} = B_{(\beta)} - \dd\Lambda_{(\alpha\beta)} ,
\end{equation}
and hence is associated to the connective structure of a
two-form $B$-field. Orthogonality between $C_+$ and $C_-$ implies that
a generic element of $X_-\in C_-$ can be written as $X_-=x+(B-g)x$.

Another way to define this structure is to introduce the
$O(2d)$-invariant generalized metric~\footnote{
  In~\cite{Gualtieri},
  the $O(d,d)$ invariant generalized metric is defined via the
  product structure $G=-{\cal J}_1 {\cal J}_2$, given two
  commuting generalized almost complex structures. This is related to
  our definition by  ${\cal H} = \eta G$.}
\begin{equation}
   {\cal H} = \left.\eta\right|_{C_+} - \left.\eta\right|_{C_-} . 
\end{equation}
Writing a general element $X=x+\xi\in E$ as $X=X_++X_-$ with
$X_\pm=x_\pm+(B\pm g)x_\pm$ one finds that the generalized metric ${\cal H}$
takes the form 
\begin{equation}
\label{eq:genmetric}
   {\cal H} = \begin{pmatrix}
         g - B g^{-1} B & B g^{-1} \\
         - g^{-1} B & g^{-1} 
      \end{pmatrix} .
\end{equation}

We can also introduce generalized vielbeins, where the local Lorentz
symmetry is replaced by $O(d)\times O(d)$. They parametrise the coset $O(d,d)/O(d) \times O(d)$ 
and encode the metric $g$ and the $B$-field. There are many different conventions one could use. 
Consider a basis of
generalized one-forms $E_A\in E^*$ with $A=1,\dots 2d$. (Note that
$\eta$ gives an isomorphism between $E$ and $E^*$ so we can equally
well think of the $E_A$ as generalized vectors.) One possibility is then to require that the metrics $\eta$ and $\mathcal{H}$ take the form 
\begin{equation}
\label{eq:O-basis}
   \eta = E^T 
      \begin{pmatrix} \id & 0 \\ 0 & -\id \end{pmatrix} E ,
      \qquad
   {\cal H} = E^T
      \begin{pmatrix} \id & 0 \\ 0 & \id \end{pmatrix} E . 
\end{equation}
Explicitly we have 
\begin{equation}
\label{eq:genvielbeind}
   E = \frac{1}{\sqrt{2}}
      \begin{pmatrix} e_+ - \hat{e}_+^T B & \hat{e}_+^T \\
         - e_- - \hat{e}_-^T B & \hat{e}_-^T 
      \end{pmatrix}
      = \frac{1}{\sqrt{2}}
      \begin{pmatrix} \hat{e}_+^T(g-B) & \hat{e}_+^T \\
         -\hat{e}_-^T(g+B) & \hat{e}_-^T 
      \end{pmatrix} \, ,
\end{equation}
where we have introduced two sets of (ordinary) vielbeins $e^a_\pm$ and
their inverse $\hat{e}_{\pm\,a}$ satisfying 
\begin{equation}
\begin{aligned}
   g &= e_\pm^T e_\pm & \text{or} &&
   g_{mn} &= e^a_{\pm\,m} e^b_{\pm\,n} \delta_{ab} \, ,\\
   g^{-1} &= \hat{e}_\pm \hat{e}_\pm^T & \text{or} &&
   g^{mn} &= \hat{e}^m_{\pm\,a} \hat{e}^n_{\pm\,b} \delta^{ab} ,
\end{aligned}
\end{equation}
and $e_\pm\hat{e}_\pm=\hat{e}_\pm e_\pm=\id$. With these conventions,
the first $d$ generalized vielbeins form a basis for $C_+$ and the
second $d$ form a basis for $C_-$. The local $O(d)\times O(d)$ action
simply rotates each set of vielbeins. Concretely we can write
\begin{equation}
\label{eq:OdOd}
   E \mapsto K E \ , \qquad
   K = \begin{pmatrix} O_+ & 0 \\ 0 & O_- \end{pmatrix} \quad 
\mbox{with} \quad O_\pm\in O(d)\, .
\end{equation}

In type II string theory compactified on a six-dimensional manifold $M$, 
the subbundles $C_\pm$ have a natural interpretation in terms of 
the world-sheet theory: they are associated to the left and right mover
sectors; $e_\pm$ are the corresponding  
vielbeins. The spinors transform under one or the other of the $O(d)$
groups. It is then usual to choose $e_+=e_-$ so that the same
spin-connections appear, for instance, in the derivatives of the two
gravitini. However, this is, of course, not strictly necessary. 

From the $O(d,d)$ action on the generalized metric and vielbein it is
straightforward to recover the familiar $O(d,d)$ transformations on
the metric, $B$-field and vielbein. 
The generalized metric (\ref{eq:genmetric}) transforms under $O(d,d)$ as
\beq
{\cal H} \rightarrow {\cal H}' = O^T {\cal H} O \, ,
\eeq
with ${\cal H}$ and $O$ given in (\ref{eq:genmetric}) and (\ref{eq:oddel}), respectively.
Given this transformation, we can derive the transformation of the bases
$e_\pm$ under $O(d,d)$. The
generalized basis forms $E^A$ transform as $E \mapsto E O$,  and
hence the vielbeins transform as 
\begin{equation} \label{etilde}
\begin{aligned}
   \hat{e}_+ &\mapsto  \left[d^T + b^T (B+g)\right]\hat{e}_+ \equiv \hat{\te}_+\\
   \hat{e}_- &\mapsto \left[d^T + b^T (B-g)\right]\hat{e}_- \equiv \hat{\te}_-  \, .
\end{aligned}
\end{equation}
This agrees with the result given in~\cite{hassan}. Note that, if we
initially set $e_+=e_-$,  generically this is no longer true after the
$O(d,d)$ transformation, and one must make a compensating Lorentz
transformation to restore the condition.  
 
It is possible to use a different set of conventions where $\eta$ and
$\mathcal{H}$ take the form  
\begin{equation}
\label{eq:Om-basis}
   \eta = {\cal E}^T 
      \begin{pmatrix} 0  & \id \\  \id & 0  \end{pmatrix} {\cal E} ,
      \qquad
   {\cal H} = {\cal E} ^T
      \begin{pmatrix} \id & 0 \\ 0 &  \id \end{pmatrix} {\cal E} . 
\end{equation}
In this basis the generalized vielbein can be written as 
\begin{equation}
\label{eq:sE}
   \mathcal{E} = \frac{1}{2}\begin{pmatrix}
             (e_++e_-) + (\hat{e}_+^T-\hat{e}_-^T)B & 
             (\hat{e}_+^T - \hat{e}_-^T) \\
             (e_+-e_-) - (\hat{e}_+^T+\hat{e}_-^T)B & 
             (\hat{e}_+^T + \hat{e}_-^T) 
          \end{pmatrix} \, .
\end{equation}
%
The $O(d)\times O(d)$ action is now of the form 
\begin{equation}
\label{eq:OdOdm}
   \cE \mapsto K \cE \ , \qquad
   K = \frac{1}{2}\begin{pmatrix} O_++O_- & O_+-O_- \\ 
      O_+-O_- & O_++O_- \end{pmatrix} \ . 
\end{equation}
As before, one can always make an $O(d)\times O(d)$ transformation to set $e_+=e_-=e$
and put the generalized vielbein into the triangular form 
\begin{equation}
\label{eq:genvielbeinm}
{\cal E} = 
\begin{pmatrix} e & 0 \\
         - \hat{e}^T B &  \hat{e}^T 
\end{pmatrix} \, .
\end{equation}
Note that in these conventions the vielbeins~\eqref{eq:sE} are
not a natural basis for $C_\pm$ since they do not diagonalise the
$O(d,d)$ metric $\eta$. However they will be of particular interest in
this paper because the latter form~\eqref{eq:genvielbeinm} is
invariant under the $\Ggeom$ subgroup of $O(d,d)$ transformations.


\subsection{$O(d,d)$ spinors} 
\label{sec:spinors}

Given the metric $\eta$, one can define $\Spin(d,d)$
spinors. These are Majorana--Weyl and we write the two helicity spin
bundles as $S^\pm(E)$. Locally, the Clifford action of $X\in E$ on
the spinors can be realized as an action on forms
$\Phi\in\left.\Lambda^{\text{even/odd}}T^*M\right|_{U_\alpha}$ given
by 
\begin{equation}
\label{cliff}
   X\cdot \Phi 
      := (x^m \check{\Gamma}_m + \xi_m \hat{\Gamma}^m)\Phi
      = i_x \Phi + \xi\wedge \Phi \, , 
\end{equation}
where  $\check \Gamma$, $\hat\Gamma$ are the $O(d,d)$ gamma matrices.
It is easy to see that 
\begin{equation}
   (XY+YX)\cdot\Phi = 2\eta(X,Y)\Phi \, ,
\end{equation}
as required. One also finds that, in going from one patch to another, the
patching of $E$ implies that 
\begin{equation} 
\label{spinorpatch}
   \Phi^\pm_{(\alpha)} 
      = \ee^{\dd\Lambda_{(\alpha\beta)}}\Phi^\pm_{(\beta)} \, ,
\end{equation}
where the exponentiated action is by wedge product. Note that the
usual action of the exterior derivative on the component forms is
compatible with this patching and gives an action 
\begin{equation}
\label{eq:dd-Dirac}
   \dd:S^\pm(E)\to S^\mp(E) \, .
\end{equation}
In terms of the  $\Spin(d,d)$ group one can view this as a
Dirac operator taking positive helicity spinors to negative
helicity spinors and \emph{vice versa}. 

Let us now return to the $\GL(d)$ action~\eqref{eq:diffeo}  on the
tangent and cotangent bundles. If we take an infinitesimal
transformation with $a=\id+\theta+\dots$, the induced action on 
the spinors is given by
\begin{equation}
   \delta\Phi = \tfrac{1}{2}\theta^m{}_n\left(
      \check{\Gamma}_m\hat{\Gamma}^n
      -\hat{\Gamma}^n\check{\Gamma}_m\right)\Phi \, .
\end{equation}
The degree of the component forms in $\Phi$ remains unchanged: in
particular, each form transforms as 
\begin{equation}
   \delta\Phi_{m_1\dots m_p} 
       = - p\, \theta^n{}_{[m_1}\Phi_{|n|m_2\dots m_p]}
           + \tfrac{1}{2}\theta^n{}_n\Phi_{m_1\dots m_p} \, .
\end{equation}
The first term correctly describes the transformation of an
element of $\Lambda^pT^*M$ under $\GL(d)$. The second
term however corresponds to a rescaling of the form by a factor of $|\det
a|^{1/2}$. This implies that  we should locally identify~\cite{Gualtieri}
\begin{equation}
\label{eq:PhiTM}
   \Phi \in |\Lambda^dT^*M|^{-1/2}\otimes 
      \left.\Lambda^{\text{even/odd}}T^*M\right|_{U_\alpha} \, .
\end{equation}

However, this presents a predicament: we cannot define the exterior
derivative on such objects, because the extra $|\Lambda^dT^*M|^{-1/2}$
factor breaks the diffeomorphism symmetry. One solution is to
identify
\begin{equation}
\label{eq:PhiL}
   \Phi \in L \otimes 
      \left.\Lambda^{\text{even/odd}}T^*M\right|_{U_\alpha} \, ,
\end{equation}
where we have introduced a new (trivial) real line bundle $L$ with
sections $\ee^{-\phi}\in L$ that transform as
\begin{equation}
   \ee^{-\phi} \mapsto  |\det a|^{1/2}\ee^{-\phi} 
\end{equation}
under the $\GL(d)$ action on $TM$, but which transform as scalars
under diffeomorphisms. We have suggestively written the sections of $L$ as
$\ee^{-\phi}$ since we will see in the next section that the
ten-dimensional dilaton indeed transforms in this way. 

Under the other two elements of $O(d,d)$ discussed in the previous section, Eqs. (\ref{eq:Btransform}) and (\ref{eq:betatransform}), 
the spinor representation transforms \beq
\label{eq:spinorodd}
\Phi^{\pm} \mapsto e^{ \omega + \beta} \Phi^{\pm} \, ,
\eeq
where $\omega$ acts by wedge product and $\beta$ by contractions.

Using the generalized vielbeins (\ref{eq:genvielbeind}) one can also introduce a basis for the
$O(d,d)$ gamma matrices $\check \Gamma, \hat\Gamma$ adapted to the
$O(d)\times O(d)$ structure. One defines 
\begin{equation}
\label{gammapm}
   \begin{pmatrix} \Gamma^+ \\ \Gamma^- \end{pmatrix}
      = (E^{-1})^T
         \begin{pmatrix} \check{\Gamma} \\ \hat{\Gamma} \end{pmatrix}
      = \begin{pmatrix}
         \hat{e}_+^T\left(\check{\Gamma} + (g - B)\hat{\Gamma}\right) \\
         \hat{e}_-^T\left(\check{\Gamma} - (g + B)\hat{\Gamma}\right) 
      \end{pmatrix} \, ,
\end{equation}
which satisfy
\begin{equation}
\begin{aligned}
   \{ \Gamma^+_a, \Gamma^-_b \} &= 0 , &
   \{ \Gamma^+_a, \Gamma^+_b \} &= 2 \delta_{ab} , &
   \{ \Gamma^-_a, \Gamma^-_b \} &= - 2 \delta_{ab} \, .
\end{aligned}
\end{equation}
One can then decompose the $\Spin(d,d)$ spinors into
$\Spin(d)\times\Spin(d)$ objects. If $d$ is even we can write
\begin{equation}
\label{eq:gammas}
   \Gamma^+_a = \gamma_a \otimes \id , \qquad 
   \Gamma^-_a = \ii\tilde{\gamma} \otimes \gamma_a \, ,
\end{equation}
where $\gamma_a$ are $\Spin(d)$ gamma matrices and
$\tilde{\gamma}=\gamma_{(d)}=\gamma_1\dots\gamma_d$ if $d/2$ is even
and $\tilde{\gamma}=-\ii\gamma_{(d)}$ if $d/2$ is odd, so that
$\tilde{\gamma}^2=\id$. Similar expressions can be written when $d$
is odd. The corresponding decompositions of the $\Spin(d,d)$ spinors are
\begin{equation}
\label{eq:Phidecomp}
   \Phi^+ = \eta_+^1 \otimes \bar{\eta}_+^2 + \eta_-^1 \otimes \bar{\eta}_-^2 \, , \qquad
   \Phi^- = \eta_+^1 \otimes \bar{\eta}_-^2 + \eta_-^1 \otimes \bar{\eta}_+^2 \, , 
\end{equation}
where $\eta_\pm^1$ and $\eta_\pm^2$ are chiral $\Spin(d)$ spinors
satisfying $\tilde{\gamma}\, \eta_\pm=\eta_\pm$. 

The generalized metric allows us to relate the $O(d)\times O(d)$
decomposition of the $O(d,d)$ spinors to the $\GL(d)$
decomposition~\eqref{eq:PhiL}. It is easiest to start by choosing the
vielbeins such that $e_+=e_-$. This identifies a common $O(d)$
subgroup of $O(d)\times O(d)$: $\eta_+^1$ and $\eta_+^2$ are now spinors of the same group
so that, under this group, $\Phi^\pm$ is a
spinor bilinear. However, any spinor bilinear can be expanded as a sum
of forms using products of gamma matrices. In particular 
\begin{equation}
\label{eq:bispinors}
\begin{aligned}
   \eta^1_+\bar{\eta}^2_+ &= \frac{1}{n_d}\sum_{\text{$p$ even}}
      \frac{1}{p!}\left(
         \bar{\eta}^2_+\gamma_{m_1\dots m_p}\eta^1_+\right)
         \gamma^{m_p\dots m_1} \, , \\
   \eta^1_+\bar{\eta}^2_- &= \frac{1}{n_d}\sum_{\text{$p$ odd}}
      \frac{1}{p!}\left(
         \bar{\eta}^2_-\gamma_{m_1\dots m_p}\eta^1_+\right)
         \gamma^{m_p\dots m_1} \, , 
\end{aligned}
\end{equation}
where $\gamma_m$ are $n_d\times n_d$ matrices, and we have
used the metric $g_{mn}$ to write the component forms in tangent space
indices. Given an expansion of the form~\eqref{eq:bispinors}, the
Clifford action on $\Phi^\pm$ is 
\begin{equation}
   X\cdot\Phi^\pm 
      = \tfrac{1}{2}[x^m\gamma_m,\Phi^\pm]_\mp
      + \tfrac{1}{2}[\xi_m\gamma^m,\Phi^\pm]_\pm \, ,
\end{equation}
where $\Phi^\pm$ are defined in~\eqref{Phidef}. 

Note, however, that the forms~\eqref{eq:bispinors} are neither twisted with
$\dd\Lambda_{\alpha\beta}$, as in~\eqref{spinorpatch}, nor transform
with the additional factor of $|\det a|^{1/2}$ under $\GL(d)$. If we
use the short-hand that $\eta^1_+\bar{\eta}^2_\pm$ represent the
corresponding sums of forms as in~\eqref{eq:bispinors}, naively we find
that the decomposition of $\Phi^\pm$ under $\GL(d)$ is 
related to the bispinor by 
\begin{equation}
   \Phi^+ = (\det g)^{-1/4}\,\ee^{-B}\eta^1_+\bar{\eta}^2_+ \, , \qquad
   \Phi^- = (\det g)^{-1/4}\,\ee^{-B}\eta^1_+\bar{\eta}^2_- \, .
\end{equation}
However, this identifies $O(d,d)$ spinors as sections
of~\eqref{eq:PhiTM}, which precludes the use of the exterior
derivative. Introducing the line bundle $L$ we can take $\Phi^\pm$ to
be sections of~\eqref{eq:PhiL}, and instead have
\begin{equation}
\label{Phidef}
   \Phi^+ = \ee^{-\phi}\ee^{-B}\eta^1_+\bar{\eta}^2_+ , \qquad
   \Phi^- = \ee^{-\phi}\ee^{-B}\eta^1_+\bar{\eta}^2_- \, ,
\end{equation}
where $\ee^{-\phi}$ is some section of $L$. By construction 
\begin{equation}
\label{eq:dilaton}
\ee^{2\phi}/\sqrt{\det g} 
\end{equation}
is invariant under $O(d,d)$. 
This is precisely the way the ten-dimensional dilaton transforms. Thus
we see that the dilaton appears very naturally in generalized
geometry: together with the generalized metric ${\cal H}$, encoding $g$ and
$B$, \emph{the dilaton defines the isomorphism between $S^\pm(E)$ and
  $\Lambda^{\text{even/odd}}T^*M$}.

Finally we note that an $O(d,d)$ spinor is said to be pure if it is annihilated by half of the gamma matrices 
(or equivalently if its annihilator is a maximally isotropic subspace of $E$). 
Any pure spinor can be represented as a wedge product of an exponentiated complex two-form with a complex $k$-form. The degree $k$ is called type of the pure spinor, and,  when the latter is closed, it serves as a convenient  way of characterizing the geometry.

A pure spinor defines an $\SU(d,d)$ structure on $E$. A further reduction of the structure group to $SU(d) \times SU(d)$ is given by the existence of a pair of compatible pure spinors.
Two pure spinors are said to be 
compatible when they have $d/2$ common annihilators.  By construction, the spinors~\eqref{Phidef} are pure and also compatible. 


\section{T-duality and $\stt$-structures}
\label{sec:Tduality}


In this section we would like to address the question of how T-duality
acts on backgrounds with  $\stt$ structure. Such geometries describe
string compactifications leading to ${\cal N}=2$ effective theories in four
dimensions and can be defined by a pair of $O(6,6)$ spinors. We shall
give explicit expressions for the new $\stt$ structure in two natural
cases. First we give the transformation of the structure when
the original manifold is a $T^3$ fibration and
we perform three T-dualities, that is the map to the mirror
configuration. Then we consider the simpler case  where the original manifold is a $T^2$ 
fibration and we perform a pair of T-dualities.
In each case, we start with a given
$\SU(3)$ structure with non-trivial $H$-flux. We shall see in
particular that T-duality can change the type of structure.  

Furthermore, in some cases we will find that naively the structure is
ill-defined. We discuss this feature in detail for the $T^2$-fibrations,
and argue that it arises precisely when the dual background is
non-geometrical. The analysis is entirely consistent with the original
discussions of non-geometry for such fibrations~\cite{Hull:2006qs,Taylor}. Here we
focus our attention on the transformation of the additional $\stt$
structure.  

Finally, we will also show that T-duality maps supersymmetric $\stt$
backgrounds to supersymmetric $\stt$ backgrounds. This requires that
the Lie derivative along the T-duality direction of the pair of
$O(6,6)$ spinors defining the geometry vanishes.  

The section begins with a general discussion of T-duality in the
context of generalized geometry. This leads to a simple expression for
the action of T-duality on $O(d,d)$ spinors which are the defining
objects for $\stt$ structures. We then review the relation between
$\stt$ structures and supergravity backgrounds, before turning to
considering T-duality on them. We conclude with the analysis of
T-duality on supersymmetric backgrounds.  

\subsection{Generalized Lie derivative, generalized Killing vectors
  and T duality}
\label{sec:localT}

In string theory T-duality is a non-local transformation. However, at
the level of supergravity, there is a corresponding transformation,
given by the Buscher rules~\cite{Buscher}, which can be viewed as a local
transformation of the supergravity fields, taking solutions to
solutions. In this section we discuss how this local T-duality
acts on the generalized structure. We will see that
formally it is simply a $O(d,d)$ gauge transformation on $E$. 

Buscher rules apply when one has a supergravity background that
admits a Killing vector field $v$ satisfying\footnote{If there is also a non-zero 
Ramond--Ramond flux,
one must further require that the Lie derivative of the flux vanishes.}
\begin{equation}
\label{eq:K-bkgd}
   \mathcal{L}_v g = \mathcal{L}_v H = 0 \, . 
\end{equation}
The condition $\mathcal{L}_vH=0$ implies that locally one can make a
gauge transformation, $B'=B+\dd\zeta'$, such that 
$\mathcal{L}_vB'=0$ or, equivalently, $\mathcal{L}_vB-\dd\zeta=0$, where
$\zeta=-i_v\dd\zeta'+\dd f$. Buscher rules are then applied to the gauge
transformed background $(g,B')$ and  generate a new background
$(\tilde{g},\tilde{B})$. Thus, in a generic gauge we require  
\begin{equation}
\label{eq:gendiff-gB}
\begin{aligned}
   \mathcal{L}_vg &= 0 \, , \\
   \mathcal{L}_vB - \dd\zeta &= 0 \, ,
\end{aligned}
\end{equation}
so that to define the T-duality action on the supergravity fields we
really need a pair $(v,\zeta)$.   

From the action~\eqref{eq:gendiff-gB} on $B$ we see that $(v,\zeta)$
act as an infinitesimal diffeomorphism generated by $v$ together with
a gauge transformation. Writing $V=v+\zeta$, we can define the
corresponding  action on sections $X=x+\xi$ of $E$ as a sort of
``generalized Lie derivative'' 
\begin{equation}
\label{eq:Lgen}
   \Lgen_V X = [v,x]_{\rm Lie} + (\mathcal{L}_v\xi - i_x\dd\zeta) \, ,
\end{equation}
where $[v,x]_{\rm Lie}$ is the Lie bracket and $\mathcal{L}_v$ is the ordinary
Lie derivative. This combination of $V$ and $X$ is actually none other
than the Dorfman bracket~\cite{Gualtieri,K-S} whose antisymmetrization gives
the Courant bracket~\eqref{eq:Courant}. It is the derived bracket for
the exterior derivative $\dd$.  

Note that this action is very natural given the bundle
structure~\eqref{eq:Edef}. We naturally identify as equivalent bundles
$E$ which are related by diffeomorphisms of the manifold $M$ and gauge
transformations which preserve the
patching~\eqref{eq:patch}. Infinitesimally, together these are
equivalent to an action of the generalized Lie derivative. 

Given this definition of $\Lgen_V$ on generalized vectors, it is then
natural to define the generalized Lie derivative of ${\cal H}$ by 
\beq
   (\Lgen_V \mathcal{H})(X,Y) = \Lgen_V\left[ {\cal H}(X,Y) \right] 
      - \mathcal{H}(\Lgen_VX,Y) - \mathcal{H}(X,\Lgen_VY) \, . 
\eeq

This is in analogy to the construction for a conventional
Lie derivative and here, when acting on a scalar function such as
$\mathcal{H}(X,Y)$, we define $\Lgen_Vf=\mathcal{L}_vf=i_v\dd f$. It is then
easy to see that 
\begin{equation}
   \Lgen_V\mathcal{H} = \begin{pmatrix}
         \begin{array}{l}
            \mathcal{L}_vg - (\mathcal{L}_vB-\dd\zeta)g^{-1}B \\
            \qquad\quad 
            - B(\mathcal{L}_vg^{-1})B - Bg^{-1}(\mathcal{L}_vB-\dd\zeta)
         \end{array} 
         & 
         (\mathcal{L}_vB-\dd\zeta)g^{-1} + B(\mathcal{L}_vg^{-1}) 
         \\*[15pt]
         -g^{-1}(\mathcal{L}_vB-\dd\zeta) - (\mathcal{L}_vg^{-1})B
         & 
         \mathcal{L}_vg^{-1}
      \end{pmatrix} \, .
\end{equation}
(Note that a similar calculation implies for the $O(d,d)$
metric~\eqref{eq:eta} that $\Lgen_V\eta=0$.) The
requirement~\eqref{eq:gendiff-gB} on $(g,B)$ then simply translates
into
\begin{equation}
\label{eq:genKilling}
   \Lgen_V G = 0 \, ,
\end{equation}
or, in other words, that $V$ defines a ``generalized Killing vector''. 

Given a generalized Killing vector $V$, we can then define the
corresponding Buscher duality as follows. First recall that there was
really an ambiguity in $V=v+\zeta$, since the generalized Lie
derivative only depends on $\dd\zeta$ so we can always shift $\zeta$ 
by $\dd f$ for an arbitrary function $f$. Using this freedom we can
always normalize $V$
\begin{equation}
   \eta(V,V) = 1 \, .
\end{equation}
Concretely, for any vector field $v$ we can introduce a coordinate $t$
such that $v=\del/\del t$. In addition, from~\eqref{eq:K-bkgd}, we know
we can write $\zeta=-i_v\dd\zeta'+\dd f$. Setting $f=t$ we have
\begin{equation}
   V = \del/\del t + (\dd t - i_{\del/\del t}\dd\zeta') \, ,
\end{equation}
and hence $\eta(V,V)=1$. We then construct the $O(d,d)$ element 
\begin{equation}
\label{eq:TVdef}
   T_V = \id - 2VV^T\eta \, .
\end{equation}
The condition $\eta(V,V)=1$ implies that
$\eta(T_VX,T_VX)=\eta(X,X)$ so $T_V\in O(d,d)$ and, in addition,
$T_V^2=\id$. 
We can choose local bases on $TM$ and $T^*M$ such that, if 
$\hat{e}_1=v=\del/\del t$ is the first basis element of $TM$ and its
dual one-form $e^1=\dd t$ is the first element for $T^*M$. Then taking
$\zeta'=0$, the T-duality matrix reads  
\beq \label{T1}
T_{\tilde{e}_1+e^1}
  =\begin{pmatrix} \id - m & m \\ m & \id - m  \end{pmatrix}  \ , \qquad 
m=\begin{pmatrix} 1 & 0 & ... & 0 \\ 0 & 0 & ... & 0 \\ \vdots & \vdots & \ddots  &\vdots \\ 0 &0 & ... & 0 \end{pmatrix} \, .
\eeq

The T-dual generalized metric $\tilde{\cal H}$ is simply given by   
\begin{equation}
   \tilde{\cal H}(X,X) = \mathcal{H}(T_VX,T_VX) \, , 
\end{equation}
or $\tilde{\cal H}=T_V^T \mathcal{H} T_V$. The action on $O(d,d)$ spinors is by an element 
of $\Pin(d,d)$ equal simply to 
the Clifford action of $V$
\beq \label{Tv}
\tilde \Phi= T_V \Phi = i_{\del/\del t} \Phi + \zeta \wedge  \Phi \, ,
\eeq   
where $\zeta=\dd t - i_{\del/\del t}\zeta'$. 

Note that T-duality is usually defined in the gauge where the NS
two-form is given by $B'=B+\dd\zeta'$ and hence satisfies
$\mathcal{L}_vB'=0$. Here we see that $T_V$ can be written as  
\begin{equation}
   T_V = \ee^{\dd\zeta'}\cdot T_{V_0} \cdot \ee^{-\dd\zeta'} , 
\end{equation}
where $V_0=\del/\del t+\dd t$. Thus the action of $T_V$ is to first
make a gauge transformation on $\mathcal{H}$ to set the
NS two form to $B'$, and then act by conventional T-duality. Note also that, as always 
for Buscher
duality, the choice of coordinate $t$ used to write $v=\del/\del t$ is
not unique. However, the effect after T-duality is simply an
additional gauge transformation. 



\subsection{$\stt$ structures and supergravity}
\label{sec:stt}

In type II supergravities compactified on a six-manifold $M$, the two
supersymmetry parameters decompose into two chiral $\Spin(6)$ spinors
transforming under the $\Spin(6)$ groups associated with $C_+$ and
$C_-$, respectively. When considering either a supersymmetric
background, or a background leading to a low-energy supersymmetric
effective action (such as a Calabi--Yau manifold with non-zero
fluxes), the supersymmetry picks out a particular pair of globally
defined, nowhere vanishing spinors $(\eta^1_+,\eta^2_+)$. 

Since $\Spin(6)\simeq\SU(4)$, a single spinor $\eta_+$ is invariant
under an $\SU(3)$ subgroup of $\Spin(6)$. If $\eta_+$ is globally
defined and nowhere vanishing, it defines an $\SU(3)$ structure. This
is a topological restriction: the tangent bundle $TM$ is
patched using only $\SU(3)$ transformations. It is equivalent to the
existence of a pair of globally defined, nowhere vanishing forms
$J\in\Lambda^2T^*M$ and $\Omega\in\Lambda^3T^*_\bbC M$. Thus the pair
of spinors $(\eta^1_+,\eta^2_+)$ defines a pair of $\SU(3)$
structures. More precisely they are invariant under an $\stt$ subgroup
of $O(6,6)$, and we say they define an $\stt$ structure. Note that the
common subgroup preserving both $\eta^1_+$ and $\eta^2_+$ is
generically $\SU(2)$, though at points where they are parallel it
becomes $\SU(3)$; in this sense  $\eta^1_+$ and $\eta^2_+$  define a ``local'' $\SU(2)$
structure. 

Thus we see that the $\stt$ structure can be defined in a number of equivalent ways:
\begin{enumerate}
\renewcommand{\labelenumi}{(\alph{enumi})}
\item the generalized metric $\mathcal{H}$ (defining $g$ and $B$) together
   with the pair $(\eta^1_+,\eta^2_+)$;
\item the two pairs of $\SU(3)$ structures $(J^+,\Omega^+)$ and
   $(J^-,\Omega^-)$ together with $B$;
\item the (local) $\SU(2)$ structure, together with a complex scalar
   $\bar{\eta}^1_+\eta^2_+$ and $B$;
\item a pair of complex generalized spinors $\Phi^\pm\in
   S^\pm_\bbC(E)$.
\end{enumerate}
The relations between these various descriptions are as
follows. First we fix the normalization of the spinors:
$\bar{\eta}^1_+\eta^1_+=\bar{\eta}^2_+\eta^2_+=1$. 
The two $SU(3)$ structures are defined as 
\bea
\label{eq:SU(3)}
&& J^+_{mn} = - i \bar{\eta}^1_+ \gamma_{m n} \eta^1_+ , \qquad  J^-_{mn} = - i \bar{\eta}^2_+ \gamma_{m n} \eta^2_+ \, , \\
&& \Omega^+_{mnp} = - i \bar{\eta}^1_- \gamma_{m np} \eta^1_+ , \quad  \Omega^-_{mnp} = - i \bar{\eta}^2_- \gamma_{m np} \eta^2_+ \, .
\eea
Here and in all the following definitions $\gamma_m$ are $\Spin(6)$ gamma matrices and
$\gamma_{(7)}=-i \gamma_1\dots\gamma_6$. These two 
$SU(3)$  structures are defined on $C_+$ and $C_-$, respectively. As such they can be always written in a standard form in terms of the vielbeins $e_\pm$
\bea
\label{eq:JOpm}
J_\pm &=& e_\pm^1 \wedge e_\pm^4 + e_\pm^2 \wedge e_\pm^5+e_\pm^3 \wedge e_\pm^6 \, ,
\nn \\
\Omega_\pm &=& (e_\pm^1+\ii e_\pm^4) \wedge (e_\pm^2+\ii e_\pm^5) 
\wedge (e_\pm^3 + \ii e_\pm^6) \, . 
\eea

Locally the two $\SU(3)$ structures define an $\SU(2)$ structure. The latter is defined by a complex one form $z=v+\ii v'$, and a triplet of real two-forms $(J_1,J_2,J_3)$, or, 
equivalently, a real two-form $j$ and a complex two-form $\omega$ \cite{waldram}. One can then always 
express the two $\SU(3)$-structures in terms of the $\SU(2)$ objects, though the 
decomposition is not unique, since it depends on the different choices of $j$ within the 
triplet $(J_1,J_2,J_3)$. Here we will use a decomposition where $j$ is
naturally associated to $(J_+,\Omega_+)$ and  $\eta^2_+= k_\| \eta^1_+
+ k_\bot (v+iv')_m \gamma^m \eta^1_-$. This  gives 
\beq
\baed
\label{eq:embed}
 J^{+} &= j - \tfrac{\ii}{2} \, z \wedge \bar z \ , & J^- & =(|k_\||^2-|k_\bot|^2) J^+ +  \mathrm{Re} (\bar{k}_{\|} k_\bot \bar \omega) - 4 \ii |k_\bot|^2 \, z \wedge \bar z \, ,\\
 \Omega^{+}& =  z \wedge \omega  \ , & \Omega^- & =k_\|^2 \Omega^+-k_\bot^2 \bar{\omega} \wedge z - 4 k_\bot k_\| j \wedge z \, .
\eaed
\eeq

To define the pure spinors we must decompose under
the two $\Spin(6)$ subgroups of $\Spin(6,6)$. We can realize the $O(6,6)$ gamma matrices as
\beq
\Gamma^+_m = \gamma_m \otimes \id \qquad \Gamma^-_m = \gamma_{(7)} \otimes \gamma_m \, .
\eeq
Here we are implicitly assuming
that $e_+=e_-$.  One can use this decomposition to write the $O(6)\times O(6)$
spinors as $\Spin(6)$ bispinors. For example, if $\eta^1_+$ and
$\eta^2_+$ are chiral spinors of the first and second $\Spin(6)$ group,
respectively, we can write 
\begin{equation}
\label{eq:Phidef}
   \Phi^+ = e^{- \phi-B} \eta_+^1\bar{\eta}_+^2 \in S^+(E), \qquad 
   \Phi^- = e^{- \phi-B} \eta_+^1\bar{\eta}_-^2 \in S^-(E) . 
\end{equation}
%
Explicitly the two pure spinors read
\bea 
\label{sttspinorsp1}
\Phi^+ &=&  e^{-\phi-B+\frac{1}{2} z \wedge \bar z} \, (\bar{k}_\| e^{-ij}  - \ii \bar{k}_\bot \omega) \ ,\\
\label{sttspinorsm1}
\Phi^- &=&   e^{  -\phi-B} \, z \, ( k_\bot e^{-ij}  + \ii \, k_\| \omega )\,  . 
\eea

\subsection{Examples}
\label{sec:examples}

In this section we first determine the 
structure of the mirror of a generic manifold with a $T^3$ fibration, by doing T-dualities 
along the $T^3$ fiber. In particular we construct explicitly the resulting
mirror local $\SU(2$) structure.   
Mirror symmetry transformations
of pure spinors for $\SU(3)$ and $\SU(2)$ structure were also studied
in \cite{PS2}, by doing  Fourier-Mukai transforms of the pure spinors.  

As discussed in the previous section, the patching of the $B$-field, (\ref{Bpatch}),
induces the patching (\ref{spinorpatch}) on the spinors so that   $e^{-B} \Phi$ is globally
well defined on $E$.  It is well known that, under a single T-duality, the 
components of the $B$-field with no legs along the T-dualized directions stay unchanged, 
while those with one leg are exchanged with the connection 
along the T-dualized fiber \cite{BEM}. In that case, spinors are still globally well-defined.
Under a second T-duality, however, if the original $B$-field has a component with both legs 
along the T-dualized directions, there is no connective structure allowing to define objects 
globally.

In the rest of this section we focus on the T-duality 
action for the latter case. We will illustrate this with 
two simple toroidal examples, where the $B$-field is purely along the two directions to be dualized.

All the calculations we perform in this Section are local. We come back to global issues in 
Section  \ref{sec:global}.

\subsubsection{Mirror symmetry on an $T^3$-fibered manifold with $H$-flux}
\label{sec:CY}

Consider the case of a manifold with a $T^3$-fibration and generic $B$ field. We assume there is an
$\SU(3)$ structure such that T-duality on the $T^3$-fibration corresponds to mirror symmetry, that 
is that the $T^3$ fibers are special-Lagrangian. We then act by T-duality and ask what is the 
structure of the mirror compactification. For the case of no $B$-field with two legs along the 
fiber, this computation was done in \cite{FMT}, where it was found that $\Phi^+$ and $\Phi^-$ are 
exchanged under mirror symmetry. Here, with generic $B$, we find that the new structure is $\stt$ 
rather than $\SU(3)$. 

We use the same notation as in \cite{FMT}, except that we denote the coordinates of $T^3$ fibration 
by $(y^1, y^2, y^3)$, and those of the base  by 
$(x^1, x^2, x^3)$. The metric and B-field are
\bea
ds^2&=&g_{ij} dx^i dx^j + h_{\alpha \beta} \eta^\alpha \eta^\beta \ ,   \nn \\
B_2&=& \frac{1}{2} B^{(0)}_{ij} dx^i \wedge dx^j + \frac12  B^{(1)}_{i \alpha} dx^i \wedge (dy^{\alpha} + 
\eta^{\alpha}) + \frac12 B^{(2)}_{\alpha \beta} \eta^{\alpha} \wedge \eta^{\beta} \, ,
\eea
where $\eta^{\alpha}\equiv  dy^{\alpha} + \lambda^\alpha_i dx^i$ and
the superindex on $B$ indicates the number of legs along the fiber.
The vielbein is $(e^{a'}_i dx^i, e^a_{\alpha} \eta^{\alpha})$, where $a,a'={1,2,3}$ are
 fiber and base orthonormal indices, respectively
\beq
\delta_{a' b'} e^{a'}_i e^{b'}_j= g_{ij} \ , \quad  \delta_{ab} e^a_{\alpha} e^b_{\beta}= h_{\alpha \beta} \, .
\eeq
The holomorphic vielbeins are
\beq
Z^a= e^a_{\alpha} \, \eta^{\alpha} + \ii \, \delta^a_{a'} \, e^{a'}_i dx^i \, .
\eeq
The original $\SU(3)$ structure is given by the pure spinors \footnote{For a generic $B$-field,
$B \wedge \Omega \neq 0$ so strictly speaking the original structure is not $\SU(3)$.}
\bea
\Phi^+&=&e^{-\phi-B-iJ} \ , \qquad 
J= \frac{\ii}{2} Z^a \bar Z^{\bar a} \, \nn \\
\Phi^-&=&e^{-\phi-B} \Omega \ , \qquad \Omega= \frac{1}{6} \epsilon_{abc} Z^a \wedge Z^b \wedge Z^c \, .
\eea
The three T-dualities on the fiber are generated by the generalized vectors $V_\alpha=\del/\del y^\alpha+d y^\alpha$. Writing $T=T_{V_1}T_{V_2}T_{V_3}$, we have $T\Phi^+=\tilde \Phi^-$, $T\Phi^-=\tilde \Phi^+$, where $\tilde \Phi^-, \tilde \Phi^+$ can be written in the form given in 
(\ref{sttspinorsp1}), (\ref{sttspinorsm1}) with
\bea
\tilde z&=&\frac{1}{|B^{(2)}|} \epsilon_{abc} B^{(2)}_{bc} \tilde Z^a \ , \qquad e^{-\tilde \phi}=e^{-\phi} \sqrt{|h+B^{(2)}|} \, , \nn \\
\tilde \jmath&=& \frac{\ii}{2} \tilde Z^a \bar{\tilde Z}^a  -\frac{\ii}{2} z \wedge \bar z\  , \qquad k_\bot =\ii |B^{(2)}| \frac{\sqrt{h}}{\sqrt{|h+B^{(2)}|}} \, , \nn \\
  \nn \\
\tilde \omega&=&-\frac{1}{|B^{(2)}|} B^{(2)}_{ab} \tilde Z^a \tilde Z^b \ , \qquad k_\| = \frac{\sqrt{h}}{\sqrt{|h+B^{(2)}|}} \, ,
\eea
where $|B^{(2)}|=\sqrt{B^{(2)}_{ab} B^{(2)}_{ab}}$, and the dual holomorphic coordinates are
\beq
\tilde Z^a=\tilde e^a_{\alpha} \tilde \eta^{\alpha} + \ii \delta^a_{a'} \tilde e^{a'}_i dx^i \ , 
\qquad  
 \  \tilde \eta^{\alpha}=dy^{\alpha} + \tilde \lambda^{\alpha}_i dx^i \, .
 \eeq
The dual (plus) vielbeins for the dual metric and the connection are related to the original ones by
\beq \label{CYdualviel}
\tilde e^{a'}_i = e^{a'}_i \ , \qquad \ \tilde e^a_{\alpha} = e^a_{\beta} \left((h+B^{(2)})^{-1}\right)^{\beta \alpha} \ , \qquad \tilde \lambda^\alpha_i=B^{(1)}_{i \alpha}
\eeq
The dual $B$-field is
\beq \label{CYdualB}
\tilde B^{(0)}=B^{(0)}  \ , \tilde B^{(1)}_{i \alpha}=\lambda^\alpha_i \ , \tilde 
B^{(2)}_{\alpha \beta}=-  \left((h+B^{(2)})^{-1}\right)^{\alpha \lambda} B_{\lambda \rho}  \left((h-B^{(2)})^{-1}\right)^{\rho \beta} \ ,
\eeq  
as expected from Buscher rules. Note that in orthonormal indices $\tilde B^{(2)}_{ab}=-B^{(2)}_{ab}$.

 In the limit of vanishing $B^{(2)}$, we recover the results of \cite{FMT}, namely $\Phi^+$ and $\Phi^-$ get exchanged under T-duality (if we write them in terms of dual vielbeins), and define
 a good mirror SU(3) structure. For nonzero $B^{(2)}$, we get a mirror $\stt$ structure that can
 be defined in patches, but does not appear to make sense globally. In the following we will focus on this issue in more detail in the slightly simpler case of a pair of T-dualities with $B$-field only on the $T^2$ fibres.


\subsubsection{Two T-dualities on $T^2$-fibration with $H$-flux}
\label{sec:T6H}

We now consider the simpler example of a $T^2$-fibered manifold with $\SU(3)$ structure defined by
\bea \label{SU30}
\Phi^+&=&e^{-\phi-B-\ii J} \ , \qquad J=e^1 \wedge e^4 + e^2 \wedge e^5+e^3 \wedge e^6 \  \qquad \nn \\
\Phi^-&=& e^{-\phi-B} \Omega \ , \qquad  \  \ \Omega=(e^1+\ii e^4) \wedge (e^2+\ii e^5) \wedge 
(e^3 + \ii e^6) \, ,
\eea
where $e^i$ are a set of vielbeins. We will also assume there is a $B$-field  along the fibre only. 
We will further assume that the fibration is trivial, implying we can introduce coordinates such 
that $e^i=r_idx^i$ etc for the fibered directions. It would be straightforward to include a 
non-trivial fibration but it is well known that this is T-dual to a non-trivial $B$-field, 
so we instead consider the latter.  

We will consider two distinct cases, where the $T^2$-fibration lies along
$e^1$ and $e^4$  and $e^2$ and $e^3$, respectively. These two cases are inequivalent with 
respect to the $\SU(3)$ structure.

\subsubsection*{Not type changing}

We consider first the case where the $T^2$ fibration lies in two directions,  $e^1$ and $e^4$,
which are paired by the complex structure.
The B-field on the $T^2$-fiber can be written as
\beq \label{B-SU30}
B= \frac{b}{r_1 r_4}  e^1 \wedge e^4 = b\, dx^1 \wedge dx^4 \ .
\eeq
The factor $b$ in the B-field can be a function of the base. For instance if the base was $T^4$, 
that is we compactify by identifying $x^i \sim x^i+1$, we could take for example $b=h x^6$, 
corresponding to a flux $H_{146}=h$ (in coordinate indices). 

We now perform two T-dualities along $\del_1$, $\del_4$, that is, in the notation of Section~\ref{sec:localT}, using the two generalized vectors $V_1=\del_1+dx^1$ and $V_4=\del_4+dx^4$.
We obtain again an $SU(3)$ structure of the form
\bea \label{SU3d}
&\tilde \Phi^+ = e^{\ii \theta_+} e^{ -\tilde \phi -\tilde B-\ii \tilde J} \ ,  &
\tilde J =\frac{e^1 \wedge e^4}{b^2+ r_1^2 r_4^2 } + e^2 \wedge e^5+e^3 \wedge e^6 \ ,  \\
&& \tilde \Omega = \frac{b-\ii r_1r_4}{b^2+r^2_1r^2_4}(\frac{r_4}{r_1} e^1+ \ii \frac{r_1}{r_4} e^4) \wedge (e^2+\ii e^5) \wedge 
(e^3 + \ii e^6) \ ,   \\
&\tilde \Phi^- = e^{\ii \theta_ -\tilde \phi -\tilde B} \tilde \Omega \ , \qquad  
 & \tilde B  = - \frac{b}{r_1r_4(b^2+ r_1^2 r_4^2)} \, e^1 \wedge e^4 \ ,\\
&&  e^{-\tilde \phi} =e^{-\phi} \sqrt{b^2+ r_1^2 r_4^2} \, ,\\
&&  {\rm tan} \theta_\pm =\mp \frac{r_1 r_4}{b} \, .
\eea
where $\tilde \Phi^{\pm}=T_{V_1} T_{V_4}(\Phi^\pm)$, are the dual pure spinors. This structure can be rewritten
using either the set of $\tilde e^i_+$ vielbeins or $\tilde e^i_-$. These read\footnote{\label{ff}The vielbeins are computed by inverting (\ref{etilde}), where we read $a, b$ from the $O(6,6)$ matrix generating the T-duality action of this example
\begin{equation*}
T_{V_1} T_{V_4}=\begin{pmatrix} a & b \\ b & a \end{pmatrix}\ , \qquad a = \begin{pmatrix} {\id}_3 - m & 0
 \\ 0 & {\id}_3 - m  \end{pmatrix} \qquad b= \begin{pmatrix} m & 0 \\ 0 & m
\end{pmatrix} \ , \quad
m \equiv  \begin{pmatrix} 1 & 0 & 0 \\ 0 & 0 & 0 \\ 0 & 0 & 0  \end{pmatrix} \ .
\end{equation*}}  
\bea \label{dv0}
\tilde e_\pm^1&= \frac{\pm r_4^2 \, e^1  - b \, \tfrac{r_1}{ r_4} e^4 }{|h+B|_f}&=r_1 \frac{\pm r_4^2 \, 
dx^1  - b \, dx^4}{|h+B|_f} \nn \\
 \te^4_\pm&=\frac{\pm r_1^2 \, e^4  + b \, \tfrac{r_4}{ r_1} e^1 }{|h+B|_f}&=r_4 \frac{\pm r_1^2 \, dx^4  
+ b \, dx^1}{|h+B|_f} \ , \\
\te^i_\pm& =e^i \, ,\   i\neq 1,4 \, ,
\eea
where $|h+B|_f= b^2+r_1^2 r_4^2 $ is the determinant of the matrix $h+B$ along the fiber directions $e^1$ 
and $e^4$. The structure after T-duality is still $\SU(3)$, since $\tilde J_+=\tilde J_-=\tilde J$, and 
$\tilde \Omega_+=\tilde \Omega_-=\tilde\Omega$.

\subsubsection*{Type changing}

We now turn to the next to simplest example. Here we assume the $T^2$-fibration lies along the $e^2$ and 
$e^3$ directions,  i.e. on two directions not paired 
by the complex structure. We again assume that $B$ lies solely along the fibration so 
that\footnote{Strictly speaking the following is not an SU(3) structure, since  $B \wedge \Omega \neq 0$. 
However, we can add for example an $e^5 \wedge e^6$ component to make $B$  proportional to 
$ \re(z^2 \wedge \bar z^3)$. Since we will perform T-dualities in $\del_2, \del_3$, the additional 
component would play no role, and stay unaffected by the T-duality.} 
\beq
 B= \frac{b}{r_2 r_3}  e^2 \wedge e^3 = b \, dx^2 \wedge dx^3 \ .
\eeq 

Performing two T-dualities generated by $V_2=\del_2+dx^2$ and $V_3=\del_3+dx^3$, we now get a local $SU(2)$ 
structure on dual space. The structure is  
defined by the pure spinors $\tilde \Phi^+=T_{V_2}T_{V_3} (\Phi^+)$, $\tilde \Phi^-= T_{V_2}T_{V_3}(\Phi^-)$,
 with $\tilde \Phi^+$, $\tilde \Phi^-$ given in (\ref{sttspinorsp1}), (\ref{sttspinorsm1}) and where the
SU(2) structure can be written in terms of $\tilde e^i_+$ 
\beq
\baed
\tilde z &=-\ii (\tilde e_+^1 + \ii \tilde e_+^4 ) \  , &  
\  \qquad
 \\
\tilde \jmath &= \tilde e_+^2 \wedge \te_+^5 + \te_+^3 \wedge \te_+^6  \ , &
k_\bot &=\ii \frac{r_2 r_3}{ \sqrt{b^2+ r_2^2 r_3^2}} \ ,  \\
\tilde \omega  &= (\te_+^2 + \ii \te_+^5) \wedge (\te_+^3 + \ii \te_+^6) \ ,  &
k_\| &=  \frac{b}{\sqrt{b^2+ r_2^2 r_3^2}} \, \\
\tilde B &=- \frac{b}{r_2 r_3}  \te_+^2 \wedge \te_+^3 \ , & e^{-\tilde \phi}&=e^{-\phi} \sqrt{b^2+ r_2^2 r_3^2} 
\eaed
\eeq
The T-dual vielbeins are\footnote{The dual vielbeins are again computed from (\ref{etilde}) where the $O(6,6)$ matrix generating the T-duality action for this case is
\begin{equation*}
T_{V_2}T_{V_3}=\begin{pmatrix} a & b \\ b & a \end{pmatrix}\ , \qquad a = \begin{pmatrix} m & 0 \\ 0 & {\id}_3 
\end{pmatrix} \qquad b= \begin{pmatrix} {\id}_3 - m & 0 \\ 0 & 0 \end{pmatrix} \ , \quad
\end{equation*} 
with $m$ taking the same form as in footnote~\ref{ff}.} 
\bea \label{dv1}
\tilde e_\pm^2&= \frac{\pm r_3^2 \, e^2  - b \, \tfrac{r_2}{ r_3} e^3 }{|h+B|_f}&=r_2 \frac{\pm r_3^2 \, dx^2  - b \, dx^3}{|h+B|_f} \nn \\
 \te^3_\pm&=\frac{\pm r_2^2 \, e^3  + b \, \tfrac{r_3}{ r_2} e^2 }{|h+B|_f}&=r_3 \frac{\pm r_2^2 \, dx^3  + b \, dx^2}{|h+B|_f} \ , \\
\te^i_\pm &=e^i \, ,\   i\neq 2,3 \, ,
\eea
and $|h+B|_f= b^2 + r_2^2 r_3^2 $.
The T-dual structure is an SU(2) since, unlike the case in the previous example, there are relative signs between $\tilde J_+$ and $\tilde J_-$:
\beq
\baed
\tilde J_\pm=& \, \te_\pm^1 \wedge \te_\pm^4 + \te_\pm^2 \wedge \te_\pm^5+\te_\pm^3 \wedge \te_\pm^6 \\
= & \, e^1 \wedge e^4+ \frac{1}{|h+B|_f} \left(\pm r_3^2 \, e^2 \wedge e^5 \pm r_2^2 \, e^3 \wedge e^6 + b  \, \tfrac{r_3}{r_2} \, e^2 \wedge e^6 - b \,  \tfrac{r_2}{r_3} \, e^3 \wedge e^5 \right) \ ,
\eaed
\eeq
and similarly for $\tilde \Omega_+$, $\tilde \Omega_-$. Because of
these relative signs, the $\stt$ structure  defined on $E$
reduces to a local $SU(2)$ on $TM$. For $b=0$, the T-dual structure
would be a ``static $SU(2)$'' ($\kpara=0$), in agreement with the examples studied in \cite{GMPT3}. The effect of $b$ in this case is to rotate this structure to a ``dynamic $SU(2)$'', with $\kpara, \kperp \neq 0$.

\subsubsection*{Relation to non-geometry}

All the discussion thus far has really been local: we have essentially
used $O(d,d)$ transformations on generalized spinors to map one local
supergravity background into another. More generally one is
interested in whether these local geometries can really be completed
into sensible global string backgrounds. It is well known that
performing T-dualities on compact backgrounds with flux can lead to
non-geometrical dual backgrounds. Non-geometry is an essentially
stringy phenomenon so we cannot expect to see it directly in the
supergravity description. In our context this relates to the fact that
T-duality does not act locally on the $T^2$ fibres. Nonetheless we see
that our examples do reflect elements of the non-geometry when one
simply takes into account that the base of the fibration is compact.  

It is well
known that a simple way of generating non-geometrical backgrounds is
to take the T-dual of a $T^2$-fibration with non-trivial $B$-field on
the fibre directions. This is precisely the case we have considered in
the previous examples.  

By construction $\Phi^\pm$ are independent of the fibre directions, as
are $\tilde\Phi^\pm$. Thus effectively one may ignore the fibre and
simply consider the dependence of the pure spinors on the base. If $b$
depends non-trivially on the base, in general, the original pure
spinors are only defined on the generalized tangent space
$E$~\eqref{eq:Edef} twisted by the one-forms $\Lambda_{(\alpha\beta)}$
encoding the non-trivial patching of the $B$ field. Put another way,
globally, in the expressions~\eqref{SU30}, the spinors $\Phi^\pm$ are
sections of $S^\pm(E)$ while $\Phi_0^+=\ee^{-\ii J}$ and
$\Phi_0^-=\Omega$ are sections of the untwisted spinor bundles
$S^\pm(TM\oplus T^*M)$.  

Now consider the T-dual pure spinors. In general we will see that they
are not well defined. That is to say, they are not sections of
$S^\pm(\tilde{E})$ for some generalized tangent bundle $\tilde{E}$ on
the dual space. This is a reflection of the fact that the dual
background is non-geometrical. To see explicitly that the spinors are
not well defined, note that they can be written in terms of the
original ones as a $\beta$-transform (\ref{eq:spinorodd})  
\beq \label{expbeta}
\tilde \Phi^\pm =  \mp \ii \, e^{-\beta}\Phi_0^\pm , \qquad
\eeq
where  the $J$ and $\Omega$ defining $\Phi_0^\pm$ take the standard
form  (\ref{SU30}) but are evaluated using the basis $e^1/r_1^2$,
$e^2$, $e^3$, $e^4/r_4^2$, $e_5$, $e_6$ for the non-type changing
example, and $e^1$, $e^2/r_2^2$, $e^3/r_3^2$, $e^4$, $e_5$, $e_6$ in
the type-changing case. The bivector $\beta$ is constructed from $B$
by changing the form indices into vector indices, namely 
\beq
\label{beta-T2}
\baed
\beta&=b \, \del/\del x^1 \wedge \del/\del x^4  &\qquad &\text {for
  non-type-changing}\ , \\ 
\beta&=b \, \del/\del x^2 \wedge \del/\del x^3  &\qquad &\text {for
  type-changing}\ . 
\eaed
\eeq

Note that this is a completely generic feature of $T^2$
fibrations. Splitting the $TM$ and $T^*M$ bundles into base and fibre
components one can write a generic $B$-transformation as the matrix 
\beq
e^B=  \begin{pmatrix} \id & 0 & 0 & 0 \\ 0 & \id& 0 & 0\\ B^{(0)} & B^{(1)} & \id & 0 \\ 
-B^{(1)} & B^{(2)} & 0 & \id \end{pmatrix} \, ,
\eeq
where the $B^{(0}$ is the component of $B$ lying solely in the base,
$B^{(1)}$ is the component with one leg in the base and one in the
fibre and $B^{(2)}$ lies solely in the fibre. If $T$ represents the
action of T-duality on the $T^2$ fibre we have 
\beq \label{TdualBtransform}
e^B\  \mapsto \ T  e^B T^{-1} 
   = \begin{pmatrix} \id & 0 & 0 & 0 \\
      -B^{(1)} & \id & 0 & B^{(2)} \\ 
      B^{(0)}& 0 & \id & B^{(1)} \\ 0 & 0
      & 0 & \id \end{pmatrix}  \ , \quad  \ 
   T=\begin{pmatrix} \id & 0 & 0 & 0 \\ 
      0 & 0& 0 & \id \\ 0 & 0 & \id & 0 \\ 
      0 & \id & 0 & 0 \end{pmatrix}  \, .
\eeq
Note that $B^{(0)}$ stays in the same position, i.e. in the T-dual
setup is still a B-transform, while $B^{(1)}$ and $B^{(2)}$ change
positions. The former plays the role of a $\GL(d)$ transformation
connecting the base and the fiber, in agreement with
(\ref{CYdualviel}), and the latter becomes a bivector
just as in~\eqref{beta-T2}.  We can
now easily understand (\ref{expbeta}). In our $T^2$ examples we took
$B^{(0)}=B^{(1)}=0$. Thus our orginal pure spinors could be written as
$e^{B^{(2)}}\Phi_0^\pm$ where $\Phi_0^+=e^{-\ii J}$ and
$\Phi_0^-=\Omega$. Then  
\beq
T( e^{-B^{(2)}} \Phi_0^{\pm})
   = T e^{-B^{(2)}} T^{-1} T \Phi_0^{\pm}
   = e^{-\beta} \tilde \Phi_0^{\pm} \ .
\eeq 
in agreement with~\eqref{expbeta}. 

We now see the basic problem. If the original $B$-field on the fibres
is non-trivial, the dual $\beta$-transform will be similarly
non-trivial. Put another way, if $\Phi^\pm$ were sections of $S^\pm(E)$
where $E$ is patched over the base by $B$-transformations along the
fibre directions, then $\tilde{\Phi}^\pm$ are sections of some bundle
were we must patch by $\beta$-transformations along the
fibres. However, this is outside the domain of conventional
generalized geometry, where, by definition $E$ can only be twisted by
$B$-transforms. Hence $\tilde\Phi^\pm$ appear to be not well defined.  

Note that in the non-type changing case the problem is even more
severe: not even the type of the pure spinors is well-defined since 
 $e^{\beta}$ changes it. The problem is not simply that the
type depends on the location in the base, but rather that one cannot
assign a unique type to the pure spinor at each point in the base. We
also note that in both cases the metric defined by the pure spinors is
similarly ill-defined, as pointed out for example in \cite{Taylor}
from Buscher rules
for T-duality \cite{Buscher}. Again, the T-dual structure makes sense
locally, but there is no good global description.     

One might have considered extending the notion of generalized tangent
space to include $\beta$-transformations. The notion of such
transforms was introduced in \cite{Gualtieri} and discussed in the
physics literature in \cite{MPZ}  in the context of supergravity duals
of deformations of conformal gauge theory, while their connection to
non-geometry was explored in \cite{PS1,PS2}. At
first sight, they seem as nice as $B$-transforms. In order to patch
the T-dual bundle, one could try and use the subgroup of $O(d,d)$
built out of $\GL(d)$ and the $\beta$-transforms defined in
(\ref{eq:betatransform}). This would correspond to identification of
$T \oplus T^*$ with an extension of $T^*$ by $T$ via
$\beta$-transform. However, unlike the $B$-transform extension this
can prove problematic. Specifically there are no consistent gluing
conditions on the two-fold overlaps that would satisfy cocycle
conditions. This can be associated with an obstruction given by the
first cohomology of the base $H^1(B, {\mathbb Z})$. We come back to this point 
in section \ref{sec:global}.


\subsection{Supersymmetric vacua and T-duality }
\label{sec:vacua}

T-duality is a powerful solution-generating tool for string theory
backgrounds. Provided the string background, that is the metric and
the fluxes, has isometries,  T-duality transformations map consistent
string backgrounds into new consistent ones. At the level of
supergravity, it maps solutions of the supergravity equations of
motion into new solutions. In this section we will show that it also maps ${\cal N}=1$ supersymmetric 
backgrounds into ${\cal N}=1$ supersymmetric backgrounds. 

The necessary conditions for
preserving ${\cal N}=1$ supersymmetry can be expressed as the 
the closure of a pure spinor, and an integrability defect of
its compatible partner which is determined by RR fluxes. Clearly, if T--duality connects two supersymmetric
backgrounds, they must separately satisfy the pure spinor
equations. 

The supersymmetry equations for ${\cal N}=1$ Minkowski vacua given in terms of the pure
spinors were found in \cite{GMPT2} and read 
\bea
\label{int}
\dd (e^{2A} \Phi_1)&=&
0 \, , \\
\dd (e^{2A}\Phi_2)&=&
 \label{nonint}   e^{2A} \dd A\wedge  \bar\Phi_2
+\frac i{16} \left[c_- e^{A} G + \ii c_+ e^{3A} *_E G \right]\ .
\eea
Let us explain the various ingredients in these equations. The pure spinors $\Phi_{1,2}$ are those of (\ref{Phidef}). The parity of $\Phi_1$ is the same as that of the RR fluxes, while $\Phi_2$ has
the opposite parity. So $\Phi_{1(2)}=\Phi_{+(-)}$ for type IIA, and the opposite for type IIB. 
$c_\pm$ are real constants that give the relation between the norm of the 10D Killing spinors and the warp factor\footnote{More precisely, $|\eta^1|^2=c_+ e^A + c_- e^{-A}$, $|\eta^2|^2=c_+ e^A - c_- e^{-A}$. Backgrounds with D-branes and/or orientifold planes require $|\eta^1|=|\eta^2|$
and therefore $c_-=0$.}, $G=e^{-B} F$ are the RR field strengths with Bianchi identity $\dd G=0$ in the absence of sources. The Hodge star  appearing in (\ref{int}) is acting on sections of $S^{\pm}(E)$. It is related to the standard one acting on $\Lambda^{\bullet} T^*$ as 
$*_E= e^{-B} * \lambda \, e^{B}$ (where $\lambda$ acts on $p$-forms by $\lambda G^{(p)}=(-1)^{[p/2]} G^{(p)}$) and it is  the chirality operator on $C_+$, as we will show later.

Note that the  ${\cal N}=1$ supersymmetry  equations are written
precisely on the objets that transform nicely under T-duality (the factor $e^{2A}$ does not play 
any role here, since as it is part of the
space-time metric, it does not transform).  
We want to claim that these equations are invariant under T-duality along a vector $v$ that preserves the background, that is 
\beq
\label{preserve}
\mathcal{L}_v \Phi_{1,2} = 0 , \qquad
\mathcal{L}_v A= 0 , \qquad 
\mathcal{L}_v G =0  \, .
\eeq
Without loss of generality we can also take the $B$-field satisfying ${\cal L}_v B=0$, so that the generalized Killing vector is $V_0=\del/\del t+dt$, with $v=\del/\del t$.  We will show that if $\Phi_{1,2}$ are a solution to the equations \eqref{int}, \eqref{nonint}, their T-duals, $\tilde \Phi_{1,2}$, solve equations of the same form, with T-dual RR field-strengths. 

We first note that 
\beq
\dd \tilde \Phi_{1,2}=\dd(\dd t \wedge  \Phi_{1,2})+ \dd (i_{\del/\del t} \Phi_{1,2})=-\dd t \wedge \dd\Phi_{1,2} -i_{\del/\del t} \dd\Phi_{1,2} + 
{\cal L}_v \Phi_{1,2}= - T_V (\dd\Phi_{1,2})
\eeq
where we have added and subtracted $i_{\del/\del t} \dd \Phi_{1,2}$ to build the Lie derivative along $v$ of
$\Phi_{1,2}$. Next, one can show that 
\beq
-T_V (\dd A \wedge \Phi_{1,2})=\dd A \wedge \tilde \Phi_{1,2} \ , \quad -T_V (G)=\tilde G  \ , \quad 
-T_V (*_E G )=\tilde{*}_E \tilde G \ .
\eeq
The first result is straightforward using~\eqref{preserve}, which implies $i_v \dd A=0$. The second one is precisely the T-duality transformation of the RR fields. Indeed, the RR fields are 
$O(d,d)$ spinors, and therefore transform as $\Phi$,  (\ref{Tv}).  The third equality needs a little more
thinking. Inserting $T^2_V=1$ we get 
\beq \label{Tlambdastar}
 -T_V *_E  T_V T_V G=(T_V *_E  T_V) \tilde G=\tilde{*}_E \tilde G \ ,
 \eeq
where in the last equality we have used that  $*_E = e^{-B} * \lambda \, e^B$ transforms by conjugation. This can be understood by noting that this combination is the chirality operator  $\Gamma^+_{(6)}$ on $C_+$. Indeed, $\Gamma^+_a$ in (\ref{gammapm}) acts on $Spin(6,6)$ spinors as
\begin{equation}
\begin{aligned}
    \Gamma^+_a \cdot G
      &= i_{\hat e_{+\,a}} G + e_{+\, a}  \wedge G - i_{\hat e_{+\, a}} B \wedge G \\
      &= \ee^{B}  (i_{\hat{e}_{+\, a}} + e_{+\,a} \wedge {} ) \, \ee^{-B} G \ ,
\end{aligned}
\end{equation} 
and therefore the chirality operator is 
\begin{equation}
  \Gamma_{(6)}^+ = 
   \frac{1}{6!}\epsilon^{a_1\dots a_6}
      \Gamma^+_{a_1}\dots\Gamma^+_{a_6}
      = \frac{1}{6!}\ee^{B}\epsilon^{a_1\dots a_6}
         (i_{\hat{e}_{a_1}} + e_{a_1} \wedge {} )
         \dots
         (i_{\hat{e}_{a_6}} + e_{a_6} \wedge {} )
         \ee^{-B}  \ ,
\end{equation}
where we have omitted the plus signs on $e$. Acting on a degree $p$ form
\begin{equation}
\begin{aligned}
   \frac{1}{6!}\epsilon^{a_1\dots a_6}
      & (i_{\hat{e}_{a_1}} + e_{a_1} \wedge {} )
         \dots
      (i_{\hat{e}_{a_6}} + e_{a_6} \wedge {} ) \, G^{(p)} \\
      &= \frac{p!(6-p)!}{6!}\epsilon^{a_1\dots a_6}
          e_{a_1}\wedge\dots\wedge e_{a_{6-p}}
          i_{\hat{e}^{a_{6-p+1}}} \dots i_{\hat{e}^{a_6}}
        G^{(p)} \\
      &= (-)^{[p/2]} * G^{(p)} = * \lambda G^{(p)} \ ,
\end{aligned}   
\end{equation}
which implies  $\Gamma_{(6)}^+=e^{B} * \lambda \, e^{-B}$. Changing
the sign of $B$ (which is  conventional, and could have been taken
opposite in (\ref{gammapm})), this is just $*_E$. Since the chirality
operator transforms under $O(d,d)$ by conjugation, we verify the last
equality in (\ref{Tlambdastar}). 

We conclude that if $\Phi_{1,2}$ are pure spinors of an ${\cal N}=1$
vacuum, their T-duals $\tilde \Phi_{1,2}$ are pure spinors of a vacuum
with T-dual RR fields.


\section{Generalized charges and the Courant bracket}
\label{sec:charges}

One of the goals of this paper is to see how aspects of non-geometry
might be encoded in the language of generalized geometry. In the
previous section we saw examples of non-geometry appearing as a result of T-duality
on backgrounds with $\stt$ structures. Specifically, assuming a
torus fibration and restricting to the base of the manifold, we saw
that the corresponding pure spinors $\Phi^\pm$ were no longer sections
of $S^\pm(E)$. Instead, to be globally defined on the base, one had to 
patch by elements of $O(d,d)$, namely a $\beta$-transform, not
contained in $\Ggeom$. 

We now turn to a related problem. It has been argued that at the level
of the effective theories non-geometric backgrounds are characterised
by certain charges, or non-geometrical fluxes, $Q$ and $R$. These are the
analogues, and T-duals, of the ``geometrical fluxes'', the $H$-flux
and the structure constants $f$ of twisted torus
compactifications. One way these fluxes appear is as structure
constants in the $2n$-dimensional Lie algebra of the effective gauged
supergravity theory~\cite{HullRE,Taylor}. These can also be derived using world-sheet Hamiltonian methods \cite{AS, NH}.

Alternatively, they appear to define a sort of generalized derivative
operator on sums of forms~\cite{Taylor2},
\begin{equation}
\label{eq:DD-def}
   \mathcal{D} = H\wedge {} + f\cdot {} + Q\cdot {} + R \llcorner \ .
\end{equation}
Here $H\in\Lambda^3T^*M$, $f\in TM\otimes \Lambda^2T^*M$, $Q\in
\Lambda^2 TM \otimes T^*M$ and $R\in \Lambda^3TM$, and the action of
$f$ and $Q$ on forms is by contraction on vector indices and
antisymmetrization on the form indices. To date these charges have
only been identified for very specific backgrounds. 

In this section we propose a generalized geometrical definition of
the generalized charges, as well as the operator $\mathcal{D}$ for generic
backgrounds. That such a formulation exists is already suggested by
the fact that $\mathcal{D}$ can be interpreted as an operator on
generalized spinors, since these are sums of odd or even forms. We
shall define the charges using the Courant bracket~\eqref{eq:Courant}
and argue that they can be interpreted as components of a generalized
spin connection~\cite{G-poisson}. A key point is that, as such, they
will be gauge dependent, taking different values depending on the
particular generalized vielbein one uses. We make the connection to
various specific examples, and discuss the global issues in the
following section. 

\subsection{The Lie bracket and the spin connection}
\label{sec:Lie}

In conventional differential geometry the Lie bracket is dual to the
exterior derivative in the sense that one can always be defined in
terms of the other. In particular, given a form $\alpha$ one has
\begin{equation}
   i_{[x,y]}\alpha = \tfrac{1}{2}\dd\left([i_x,i_y]\alpha\right)
      + i_x\dd(i_y\alpha) - i_y\dd(i_x\alpha)
      + \tfrac{1}{2}[i_x,i_y]\dd\alpha \ . 
\end{equation}
This relation implies there are two equivalent ways of defining the
spin-connection. Given any frame $e^a$ and its inverse $\hat{e}_a$ we
can define the objects $f^a{}_{bc}$ in two different ways
\begin{equation}
\label{eq:sp-conn}
 [\hat{e}_a,\hat{e}_b] = f^c{}_{ab} \hat{e}_c  \ , 
   \qquad \Leftrightarrow \qquad
\dd e^a = - \tfrac{1}{2} f^a{}_{bc}e^b \wedge e^c  \ . 
\end{equation}
If $e^a$ are vielbeins for some metric, then the requirement that the
Levi--Civita connection is metric compatible and torsion-free implies
that we can define the spin connection in terms of $f^a{}_{bc}$ as 
\begin{equation}
   \omega_{ab} 
      = \tfrac{1}{2} \left( f_{cab} + f_{acb}- f_{bca} \right) e^c \ ,  
\end{equation}
where we have raised and lowered frame indices with frame metric
$\delta_{ab}$.

\subsection{Generalized charges, brackets and a generalized spin connection} 
\label{sec:genalg}

The expression for $\omega^a{}_b$ in terms of the Lie bracket,
suggests that, in generalized geometry,  one can use the Courant bracket~\eqref{eq:Courant}  
to define a generalized spin connection
$\gO$~\cite{G-poisson}. Suppose we have a basis given by 
the generalized vectors $\mathcal{E}_A$ with $A=1,\dots,2d$, and we
use the conventions where $\eta$ and $\mathcal{H}$ take the
form~\eqref{eq:Om-basis}, or equivalently 
\begin{equation}
\label{eq:bases} 
\begin{aligned}
   \eta &= \tfrac{1}{2} \left( 
       \mathcal{E}_a \otimes \mathcal{E}^a
       + \mathcal{E}^a \otimes \mathcal{E}_a \right) \\
   \mathcal{H} &= \tfrac{1}{2} \left( 
       \delta^{ab}\mathcal{E}_a \otimes \mathcal{E}_b 
       + \delta_{ab}\mathcal{E}^a \otimes \mathcal{E}^b \right) .
\end{aligned}
\end{equation}
Here we have split $\mathcal{E}_A=(\mathcal{E}_a,\mathcal{E}^a)$ with
$a=1,\dots,d$. In the language of ref.~\cite{G-poisson} this has given
us a split of the generalized tangent space $E=C_0+C_0^\perp$ spanned
by $\mathcal{E}_a$ and $\mathcal{E}^a$ respectively. It requires that
the resulting maps from $C_0$ and $C_0^\perp$ to $TM$ and $T^*M$ are
non-degenerate. 

In analogy to~\eqref{eq:sp-conn} one can then define
\begin{equation}
   [ \cE_A, \cE_B ] = F^C{}_{AB} \cE_C .
\end{equation}
Our claim is that the components of $F^A{}_{BC}$ are the
generalized fluxes $f$, $H$, $Q$ and $R$. To see how this might work, let
us first consider some special cases. If $e^+=e^-$, the generalized
vielbeins can be written as~\eqref{eq:genvielbeinm} so that 
\begin{equation}
\label{eq:B-basis}
   \cE^a = e^a \ , \qquad
   \cE_a = \hat{e}_a - i_{\hat{e}_a}B \ . 
\end{equation}
It is then easy to calculate 
\begin{equation}
\label{eq:B-alg}
\begin{aligned}
   {}[\cE_a, \cE_b] &= f^c{}_{ab} \cE_c - H_{abc} \cE^c  \ , \\
   [\cE_a, \cE^b] &= - f^b{}_{ac} \cE^c \ ,  \\
   [\cE^a, \cE^b] &= 0 \ ,
\end{aligned}
\end{equation}
where $f^a{}_{bc}$ is defined as in~\eqref{eq:sp-conn} 
and
\beq
H_{abc} = - 3 (i_{\hat{e}_{[c}}\dd B_{ab]} + f^d{}_{[ab} B_{c]d} ) \, .
\eeq

One could also choose a basis based on the
$\beta$-transform~\eqref{eq:betatransform} where 
\begin{equation}
\label{eq:beta-basis}
   \cE^a = \tilde{e}^a + \beta\cdot \tilde{e}^a\ , \qquad
   \cE_a = \hat{\tilde{e}}_a \, ,
\end{equation}
and, in order to reproduce the generalized
metric~\eqref{eq:genmetric}, we have $\tilde{e}^a \tilde{e}^b \delta_{ab}=\tilde{g}$
\begin{equation}
\begin{aligned}
   \tilde{g} &= g - Bg^{-1} B \ , \\
   \beta &= - \tilde{g}^{-1} B g^{-1} \ . 
\end{aligned}
\end{equation}
One then finds that 
\begin{equation}
\label{eq:beta-alg}
\begin{aligned}
   {}[\cE_a, \cE_b] &= f^c{}_{ab} \cE_c \ , \\
   [\cE_a, \cE^b] &= - f^b{}_{ac} \cE^c + Q^{bc}{}_a \cE_c  \ ,  \\
   [\cE^a, \cE^b] &= Q^{ab}{}_c\cE^c + R^{abc} \cE_c \ ,
\end{aligned}
\end{equation}
where $f^a{}_{bc}$ is defined as in~\eqref{eq:sp-conn} but using
$\tilde{e}^a$ and 
\beq \label{qgen}
   Q^{ab}{}_c= i_{\hat{\tilde{e}}_c} \dd\beta^{ab} 
      + \beta^{ad} f^b{}_{cd} - \beta^{bd} f^a{}_{cd} \ , 
\eeq
and
\begin{equation} \label{Rgen}
   R^{abc} = \beta^{ad} i_{\hat{\tilde{e}}_d}\dd\beta^{bc} 
      - \beta^{bd} i_{\hat{\tilde{e}}_d}\dd\beta^{ac} 
      + \beta^{ad} \beta^{be} f^c{}_{de} \ , 
\end{equation}
where
$\beta=\tfrac{1}{2}\beta^{ab}\hat{\tilde{e}}_a\wedge\hat{\tilde{e}}_b$.
Note that the new terms in the algebra only vanish if $\beta=0$, 
showing that in contrast to a
closed $B$-transform, a constant $\beta$-transform is not an
automorphism of the Courant bracket.  For special cases though, the contractions of $\beta$ on $f$ appearing in (\ref{qgen}) and (\ref{Rgen}) vanish, as we will show in section \ref{sec:fibrations}. 

The general case is as follows. First note that $\cE_{(0)}^m=\dd x^m$
and $\cE_{(0)m}=\del_m$ is a (local) frame for the $O(d,d)$ metric
$\eta$, but not in general for $\mathcal{H}$. (Note that in this
coordinate frame $F=0$.) This implies that any given frame $\cE$ can
always be written as an $O(d,d)$ rotation of $\cE_{(0)}$, that is
$\cE =\cE_{(0)} O$ or in components 
\begin{equation}
   \cE^a = \aaa^a{}_m \dd x^m + \bbb^{am} \del_m \ , \qquad
   \cE_a = \ccc_{am} \dd x^m + \ddd_a{}^m \del_m \ . 
\end{equation}
The splitting condition implies that $\aaa^a{}_m$ and $\ddd_a{}^m$ are
non-degenerate. This then leads to the general algebra
\begin{equation}
\label{eq:gen-alg}
\begin{aligned}
   {}[\cE_a, \cE_b] &= f^c{}_{ab} \cE_c  - H_{abc} \cE^c \ , \\
   [\cE_a, \cE^b] &= - \tilde{f}^b{}_{ac} \cE^c + Q^{bc}{}_a \cE_c  \ , \\
   [\cE^a, \cE^b] &= \tilde{Q}^{ab}{}_c\cE^c + R^{abc} \cE_c \ .
\end{aligned}
\end{equation}
where the fluxes $f$, $H$, $Q$ etc are given in terms of derivatives
of $O^A{}_B$.

The commutators~\eqref{eq:B-alg} and~\eqref{eq:beta-alg}
agree in form with those appearing in the effective gauged supergravity
theories. However, there the elements $\mathcal{E}_A$ are symmetry
generators rather than generalized vectors. In addition, it is clear
that our definition of $F^A{}_{BC}$ is \emph{gauge dependent}. The
frame $\mathcal{E}_A$ is not uniquely defined; instead equivalent
frames are related by $O(d)\times O(d)$
transformations~\eqref{eq:OdOdm}. Changing 
frame thus changes the charges $f$, $H$, $Q$ and $R$. A very explicit
example is provided by the two frames~\eqref{eq:B-basis}
and~\eqref{eq:beta-basis}. Both define the same generalized metric,
but lead to very different charges~\eqref{eq:B-alg}
and~\eqref{eq:beta-alg}. In fact there is a stronger
statement. Locally, one can \emph{always} make an $O(d)\times O(d)$
transformation to go the basis~\eqref{eq:B-basis}. Thus it would
appear that locally the $Q$ and $R$ charges can always be gauged
away. As such it would seem hard, locally, to decide when a
given set of charges implies we have a non-geometrical background and
when not. We return to these points in the next section.

We want now to  make the connection between the $F^A{}_{BC}$ and the
generalized derivative $\mathcal{D}$ given in~\eqref{eq:DD-def}. 
Hitchin~\cite{Hitchin-bracket} has noted that, in analogy to the
duality between the Lie bracket and the exterior derivative, the
Courant bracket is dual to the action~\eqref{eq:dd-Dirac} of exterior
derivative on $S^\pm(E)$. Explicitly, if $X\cdot\Phi$ is the Clifford
action of $X\in E$ on an spinor $\Phi$, then  
\begin{equation}
\label{eq:Courant-d}
\begin{aligned}
   {}[X,Y]\cdot \Phi 
      &= \tfrac{1}{2}\dd\left[
            \left(X\cdot Y - Y\cdot X\right)\cdot \Phi\right]
         \\ & \qquad 
         + X\cdot \dd\left(Y\cdot \Phi\right)
         - Y\cdot \dd\left(X\cdot \Phi\right)
         + \tfrac{1}{2}\left(X\cdot Y - Y\cdot X\right)\cdot \dd\Phi 
         \ . 
\end{aligned}
\end{equation}
This suggests that the charges $F^A{}_{BC}$ can be equally well
defined using $O(d,d)$ spinors. To see how this works we need to
consider what we mean by a generalized connection. Given a tensor
bundle $W$, the ordinary Levi--Civita connection $\nabla=\del+\omega$
is differential operator $\nabla:C^\infty(W) \to C^\infty(TM^*\otimes
W)$. By analogy~\cite{G-poisson} a generalized connection is an
operator\footnote{Note that we can use the metric $\eta$ to identify
  $E$ and $E^*$}  
\begin{equation}
   D : C^\infty(W) \to C^\infty(E\otimes W)  \, ,
\end{equation}
where $W$ is some vector bundle which carries a representation of
$O(d,d)$. Again we can think of $D$ as $D=\del+\gO$, where the
ordinary derivative $\del$ simply gives a term in the $T^*M$ part of
$E$ and nothing in the $TM$ part. Thus one defines the derivative
$D$, acting on a generalized vector $X=X^A\cE_A$, as 
\begin{equation}
   D X = \left(\dd X^A + \gO^A{}_B X^B \right) \otimes \cE_A \ .
\end{equation}
Given a generalized connection one can then ask if it is compatible
with $\eta$ or with the generalized metric $\mathcal{H}$, that is
$D\eta=0$ or $D\mathcal{H}=0$, and if it is torsion free in a
generalized sense. For instance, in~\cite{Hitchin-bracket,G-poisson} a
natural $\eta$ and $\mathcal{H}$ compatible connection is defined, which is not
torsion free. If in particular one has a generalized connection that
preserves the metric $\eta$, one can define a derivative of
$\Spin(d,d)$ spinors by using the gamma matrices $\Gamma^A$ associated
to a particular frame $\cE_A$, that is   
\begin{equation}
   D_A \Phi = \del_A \Phi + \tfrac{1}{4}\gO_A{}^{BC}\Gamma_{BC}\Phi \ . 
\end{equation}

If we now return to the exterior derivative we recall that it acts as
a Dirac operator on the $\Spin(d,d)$ spinors,  $\dd:S^\pm(E)\to
S^\mp(E)$. In the particular basis~\eqref{cliff}, where the
generalized vielbein takes the form $(\cE_{(0)}^m,\cE_{(0)m})=(\dd
x^m,\del_m)$ we can write the exterior derivative in terms of a generalized
$\eta$-compatible connection $D$
\begin{equation}
   \DD \Phi = \Gamma^A D_A \Phi 
      = \hat{\Gamma}^m \del_m \Phi = \dd\Phi \  . 
\end{equation}
In this basis the spin-connection $\gO$ vanishes, consistent with the
fact that $F^A{}_{BC}=0$. As we commented above, a general frame,
which is also a basis for $\mathcal{H}$, can be written as an $O(d,d)$
rotation of $\mathcal{E}_{(0)}$. In this basis $\gO$ is non-zero and
the Dirac operator can be written as  
\begin{equation}
   \DD \Phi = O \dd \left( O^{-1} \Phi \right)
     = \dd \Phi + \left( O \dd O^{-1} \right) \Phi
\end{equation}
where by construction, we are writing the spinor $\Phi$ in a frame
associated to $\cE_A$, that is, where it can be written as a tensor
product of two spinors as in~\eqref{eq:Phidecomp}. Thus, for instance,
if $\cE_A$ takes the form~\eqref{eq:genvielbeinm}, then we write
elements of $\Phi$ in terms of the $e^a$ basis\footnote{For simplicity
  we ignore the subtleties associated to the dilaton here.}
\begin{equation}
   \Phi = e^{-B}\sum_{n=0}^d \frac{1}{n!} \Phi_{a_1\dots a_n} \, 
      e^{a_1}\wedge \dots e^{e_n}
\end{equation}
and in frame indices 
\begin{equation}
   (\DD \Phi)_{a_1\dots a_n} = n \del_{[a_1} \Phi_{a_2\dots a_n]}
      + n f^b{}_{[a_1 a_2} \Phi_{|b|a_3\dots a_n]} 
      - \frac{n!}{3!(n-3)!} H_{[a_1 a_2 a_3} \Phi_{a_4\dots a_n]} .
\end{equation}
We see the appearance of the generalized fluxes $f$ and $H$ in the
definition of $\DD$ just as in the definition of $\mathcal{D}$ given
in~\eqref{eq:DD-def}. In a more general basis one would also generate
$Q$ and $R$ terms. This is reflecting the duality between the exterior
derivative and the Courant bracket. In summary, we see that the
derivative $\mathcal{D}$ is simply the exterior derivative written in
a frame adapted to the generalized vielbein $\mathcal{E}_A$. 


\section{Global properties, generalized charges and non-geometricity}
\label{sec:global}

In the previous section we proposed a generalized geometric
expression for the charges $f$, $H$, $Q$ and $R$, which  arises
from the Courant bracket between generalized vielbeins
$\mathcal{E}_A$. This was a purely local notion. Crucially, it is also
gauge dependent: changing the frame $\mathcal{E}_A$ changes the
charges. A clear example  was provided by the two bases~\eqref{eq:B-basis}
and~\eqref{eq:beta-basis}. In fact, locally one can \emph{always}
choose the gauge where $\cE_A$ takes the form~\eqref{eq:B-basis} for
which only the geometrical $f$ and $H$ charges appear. This implies
that if $Q$ and $R$ are going to encode non-geometry, we can only see
this \emph{globally}: there must be some global obstructions to
gauging them away. 

In this section we try to address this issue in some particular
cases. We will focus on backgrounds which admit a $\mathbb{T}^d$
action. We  give first the general analysis and then focus on two
 known examples which can lead to non-geometry. The advantage
of such backgrounds is that the local fibration structure picks out a
preferred frame $\cE_A$ with respect to which one can define the
charges, which allows us to see how non-geometry can be characterized
in terms of Courant brackets. Of course, for non-geometrical
backgrounds, the $\mathbb{T}^d$ fibration will not patch to form a
proper manifold. As such it cannot be described using
supergravity. Nonetheless we will see that the twisting of the frame
over the base of the fibration can be used to 
characterize the fact that the background is non-geometrical. 

The existence of a preferred frame generically implies an additional
structure beyond $O(d,d)$. The extreme case of this are ``generalized
parallelizable'' backgrounds, where, in analogy with conventional
parallelizable manifolds, there is a globally preferred frame
$\cE_A$. We end the section with a brief generic discussion of such
backgrounds with additional structure. 


\subsection{Generalized charges and fibrations}
\label{sec:fibrations}

In Section \ref{sec:genalg} we showed how the two different choices of
bases for the generalized vielbeins, \eqref{eq:B-basis}
and~\eqref{eq:beta-basis}, give rise to two different algebrae with
charges $f$ and $H$, and $Q$ and $R$, respectively. Here we consider a
particular realisation of these two bases. More precisely we consider
a class of manifolds which are  $\mathbb{T}^d$ fibrations. The
structure of the metric and $B$-field is the same as in Section
\ref{sec:CY}, but now the dimension of the fibre is $d$ rather than
three. 

If the metric admits a $\mathbb{T}^d$ action, the generalized vielbeins can be 
written as 
\begin{equation}
\label{eq:genvfibre0}
\begin{pmatrix} \cE^a \\   \cE_a \end{pmatrix} 
= \begin{pmatrix}
e^{a'}{}_{i} & 0 & 0 & 0   \\ 
\lambda^{a}{}_i & e^{a}{}_{\alpha} & 0 & 0 \\  
B_{a' i}  &  B_{a' \alpha}  &  {\hat e}_{a'}{}^{i} & 
\hat{\lambda}_{a'}{}^{\alpha}  \\ 
B_{a i}  &  B_{a \alpha} & 0 & \hat{e}_{a}{}^{\alpha}
 \end{pmatrix}  \, \begin{pmatrix} dx^{i}\\ dy^{\alpha} 
\\  \partial_i \\\partial_{\alpha} \end{pmatrix}  \, .
\end{equation}
In order not to clutter the expression above we defined the connections
$\lambda^{a}{}_i = e^{a}{}_{\alpha} \lambda^{\alpha}{}_i$ and 
$\hat{\lambda}_{a'}{}^{\alpha} = - \hat{e}_{a'}{}^{i} \lambda_i{}^{\alpha}$.
Similarly the components of the $B$-field are
\bea 
&&  B_{a' \alpha} = {\hat e}_{a'}{}^{i} B_{i \alpha} \qquad \qquad
B_{a' i} = {\hat e}_{a'}{}^{j}  (- B_{ij} + B_{j \alpha} \lambda^{\alpha}{}_i
- \lambda_j{}^{\alpha} B_{\alpha i}) \, ,\\
&& B_{a \alpha} =   {\hat e}_{a}{}^{\beta} B_{\beta \alpha } \qquad \qquad
B_{a i}  =  - {\hat e}_{a}{}^{\alpha} ( B_{\alpha \beta} \lambda^{\beta}{}_i + B_{\alpha i}) \, .
\eea
As we see, the vectors on the base are shifted by the 
derivatives along the torus due to the nontrivial fibration. 
It is straightforward to check that the generalized vielbeins \eqref{eq:genvfibre0} satisfy 
the algebra \eqref{eq:B-alg}, where, because of the isometries in
the fibre directions, the only non trivial components of 
$f$ are  $ f^{a'}{}_{b'c'} =  i_{\hat{e}_{b'}} i_{\hat{e}_{c'}} {\rm d} e^{a'}$ and 
$ f^{a}{}_{b'c'} = - \hat{e}_{[b'}{}^i  \hat{e}_{c']}{}^j \del_i \lambda^a_j$.

As we will see in the examples below, there are at least two different ways
to obtain the $\beta$-transformed basis  by $O(d,d)$ transformations.
One possibility is to consider the torus fibration with a $B$-field with components
in the fibre direction only, and to apply T-duality along the fibre. Since the torus 
directions are isometries, this is a perfectly lecit transformation.
Alternatively we can set the $B$-field to zero and 
perform a $\beta$-deformation on the metric respecting the torus action.
In both cases the resulting vielbein has the form
\beq
\label{eq:genvfibrebeta}
\begin{pmatrix} \tilde{\cE}^a \\   \tilde{\cE}_a \end{pmatrix} 
= \begin{pmatrix} e^{a'}_{i} & 0 & 0 & 0   \\ 
\lambda^{a}_i & 1 & 0  & \beta^{a \alpha} \\ 0 & 0& {\hat e}_{a'}{}^{i} & 
 \hat{\lambda}_{a'}{}^{\alpha} \\ 0 & 0 & 0  & 1
 \end{pmatrix}  \, \begin{pmatrix} dx^{i}\\ dy^{\alpha} 
\\  \partial_i \\\partial_{\alpha} \end{pmatrix} \, ,
\eeq
with $\beta^{a \alpha} = e^a{}_{\beta} \beta^{\beta \alpha}$.
This generalized vielbein gives the algebra
\eqref{eq:beta-alg}.  Note
that the derivatives along the fiber coordinates as well as 
the contractions of $\beta$ and $f^a{}_{b'c'} $ vanish.  
Moreover 
 the  algebra (\ref{eq:beta-alg}) takes the canonical form, with $R^{abc}=0$ and the only non-vanishing component of $Q$-charge being
$Q^{ab}{}_{c'} = i_{\hat{\tilde{e}}_{c'}} \dd\beta^{ab} = \partial_{c'} \beta^{ab}$.

Note that a corollary of the above computation is that,  on a
manifold that admits a ${\mathbb T}^d$ action,
a constant $\beta$-transform  with components only 
along the ${\mathbb T}^d$ fibre is a symmetry of the Courant bracket.

On the other side, it is not hard to see that when $\beta$ lies along the
fibers, the $R$-charge is non-vanishing only if $\beta$ depends on the torus
coordinates. In our context, such a situation can arise  when the $B$-field
does not respect the isometries of the background.

As we already discussed, it is always possible to perform a local 
$O(d) \times O(d)$ transformation, \eqref{eq:OdOdm}, which preserves the 
form of the generalized metric ${\cal H}$, (\ref{eq:genmetric}), and maps the 
$\beta$-transformed basis into the usual ($B$-transformed) basis on $E$
\beq
K \,  \tilde{\cE} =  \frac{1}{2} \begin{pmatrix} O_+ + O_- &   O_+ - O_-   \\ 
O_+ - O_-  &  O_+ + O_- 
 \end{pmatrix} \,  \begin{pmatrix}  e_{\rm B} & & &   \\ 
& e_{\rm F} & & e_{\rm F} \beta \\   & & {\hat e}^T_{\rm B} &   \\ & & &   {\hat e}^T_{\rm F}
 \end{pmatrix}  =  \begin{pmatrix} e_{\rm B} & & &   \\ 
& \te_{\rm F} & &  \\   & & \hat{e}^T_{\rm B} &   \\ & - \hat{\te}^T_{\rm F} B & 
& \hat{\te}^T_{\rm F} \, .
 \end{pmatrix} 
 \eeq
where the explicit expression for the matrices $O_{\pm}$ is 
\beq
\label{eq:betacomp}
O_+ = \mathbb{I} \qquad \qquad 
O_- = \begin{pmatrix} 
1 & \\
& (\hat{e}^T_{\rm F} + e_{\rm F} \beta) (\hat{e}^T_{\rm F} - e_{\rm F} \beta)^{-1}
\end{pmatrix} \, .
\eeq

Let us examine the global issues associated with such a transformation. As explained 
earlier (see (\ref{Bpatch})) the $B$-field is defined only locally. Moreover
$B_{\alpha \beta}$ must 
not be a single-valued function in order for the  $H$-flux to be non-trivial in 
cohomology. This in turn means that the matrix $O$ and the resulting generalized 
vielbeins are not single-valued either. As a consequence the transformation in question, 
while not changing the generalized metric locally, cannot produce a well-defined metric.
Put differently, $Q^{ab}{}_{c'} $ and $H_{c'ab} $ can be deformed into each other by using local 
diffeomorphisms, 
provided they are exact. The difference in the vertical components (the position of the $a,b$ indices) is not important here - the obstruction is given by the first cohomology of the 
base of the torus fibration: when the first cohomology of the base is trivial, there simply 
do not exist any $B_{ab}$ which are not single-valued. This agrees with the T-duality obstruction derived form the world-sheet perspective \cite{Hull:2006qs, BHM}.

This is a general feature of the algebrae obtained from the generalized
vielbeins: the  $Q$ charges can be gauged away and the algebra can be smoothly deformed into 
a conventional one with $H$ and $f$ (\ref{eq:B-alg}), only
if the first cohomology of the base is trivial.


\subsection{Examples}
\label{sec:Q-examples}

In this section we illustrate with two basic and well known examples the general discussion
above. The first one is probably the simplest and best known example of non-geometric
background, namely the T-dual of the three-torus with a $B$-field along the T-duality 
directions. In this case the base of the fibration is not simply connected and we will see
that the local transformation that should gauge the $Q$-charges away does not make sense
globally.

The second example is the Lunin-Maldacena solution~\cite{LM}. This is a good geometric 
background obtained via $\beta$-transformation along the directions of the $T^2$ fiber. 
In this case we will see that the $Q$-charges can indeed be gauged away by a good 
$O(d)\times O(d)$ transformation.

\subsubsection*{Three torus with $H$-flux}

In this subsection we shall illustrate the construction on the
prototypical example of a non-geometric background: the T-dual of the
straight three-torus $T^3$ (${\rm Vol}(T^3) = dx^1 \wedge dx^2 \wedge
dx^3$) with a non-trivial NS three-form flux, $H = k dx^1 \wedge dx^2
\wedge dx^3$. As we will see,  this is an example of a general
parallelizable manifold.   

Clearly there is a basis of well-defined vectors
$\{\partial_1, \partial_2, \partial_3\}$ and a basis of one-forms
$dx^1, dx^2, dx^3$ on the tangent and the cotangent bundle,
respectively.  We choose a gauge where $B= k x^1 {\rm d}x^2 \wedge {\rm d}x^3$.
It is not hard to see that a global basis for the
sections on $E$ is given by   
\begin{equation}
\label{eq:T3E}
   (\cE_a, \cE^a) = 
     (\partial_1, \partial_2 - k x^1 dx^3, \partial_3 + k x^1 dx^2; 
     dx^1, dx^2, dx^3) \, . 
\end{equation}
Note that this is of the standard triangular
form~\eqref{eq:genvielbeinm}. Calculating the Courant bracket yields
the familiar algebra 
\begin{equation}
\label{eq:H-alg}
\begin{aligned}
   {}[\cE_a, \cE_b] &= - H_{abc} \cE^c \ , \\
   [\cE_a, \cE^b] &= 0 \ , & 
   \qquad \text{where} \quad H_{123} &= k \ , \\
   [\cE^a, \cE^b] &= 0 \ .
\end{aligned}
\end{equation}

One can now act on the basis by an element of $O(3,3)$ to go to the
T-dual configuration. T duality in the direction $x^3$  amounts to
$\partial_3 \leftrightarrow dx^3$. In order for  the new basis to be split
(that is for the projections from $C_0$ and $C_0^\perp$ to $TM$ and $T^*M$
to be non-degenerate) we have to perform a local $O(d)\times O(d)$ transformation
of the same form as the T-duality one. In this case this ends up in a relabeling of
the vielbeins. We then arrive at the basis 
\begin{equation}
   (\tcE_a, \tcE^b) 
     = (\partial_1, \partial_2 - k x^1 \partial_3, \partial_3; 
     dx^1, dx^2, dx^3- k x^1 dx^2) \, ,
\end{equation}
where we have suppressed the tildes on the dual coordinates in order
not to clutter the notation. Again this takes the standard
form~\eqref{eq:genvielbeinm}. It is well known that the dual background
is a twisted torus with zero $B$-field. This is reflected in the fact
that the new basis consists of well-defined sections of $T$ and
$T^*$. Computing the Courant bracket gives simply 
\begin{equation}
\label{ex-two}
\begin{aligned}
   {}[\tcE_a, \tcE_b] &= f^c{}_{ab} \tcE_c  \ , \\
   [\tcE_a, \tcE^b] &= -f^b{}_{ac} \tcE^c  \ , & 
   \qquad \text{where} \quad f^3{}_{12} &= k  \, , \\
   [\tcE^a, \tcE^b] &= 0 \ .
\end{aligned}
\end{equation}
where we recognize the nilpotent Heisenberg algebra given by the
structure constants $(0,0,12)$. 

The second T--duality -- now in direction $x^2$ --  acts very much the
same way and amounts to $\partial_2 \leftrightarrow dx^2$. The new
basis  (again after some relabeling) is
\begin{equation}
\label{eq:ttE}
   (\ttcE_a, \ttcE^a) 
   = (\partial_1, \partial_2, \partial_3; 
   dx^1, dx^2 + kx^1 \partial_3, dx^3- k x^1 \partial_2)    
\end{equation}
and yields an algebra
\begin{equation}
\label{eq:q-alg}
\begin{aligned}
   {}[\ttcE_a, \ttcE_b] &= 0 \ , \\
   [\ttcE_a, \ttcE^b] &= -Q^{bc}{}_a \ttcE_c  \ , & 
   \qquad \text{where} \quad Q^{23}{}_1 &= k \ , \\
   [\ttcE^a, \ttcE^b] &= Q^{ab}{}_c \ttcE^c  \ .
\end{aligned}
\end{equation}
We now note that the new basis is \emph{not} in the standard
form~\eqref{eq:genvielbeinm}. In fact, it is not a
section of $E$ for any choice of extension $T^*M\to E\to TM$. Rather,
it is an extension of $TM$ over $T^*M$. This is reflected in the fact
that the generalized metric $\mathcal{H}$ built from $\ttcE_A$ is not single
valued as a function of $x^1$. We are used to this happening because
$B$ is not single valued, but here the monodromy in $\mathcal{H}$ is a
$\beta$-transformation rather than a $B$-transformation.    

One can, of course, find a local map $O(d)\times O(d)$ map to put the
basis~\eqref{eq:ttE} into the standard
form~\eqref{eq:genvielbeinm}. Explicitly, in~\eqref{eq:OdOdm} one takes
\beq
O_+ = \mathbb{I} \ \,  \qquad \qquad  O_- = \begin{pmatrix}
         1 & 0 & 0 \\ 
         0 & \Delta^{-2} (1 - k^2 x^2) & - \Delta^{-2} 2 k x^1 \\
         0 & \Delta^{-2}2 k x^1 & \Delta^{-2} (1 - k^2 x^2)
   \end{pmatrix} \ ,
\eeq

%
where $\Delta=\sqrt{(kx^1)^2+1}$. The new basis is then 
\begin{equation}
   (K\ttcE_a, K\ttcE^a) 
   = (\partial_1, \Delta\partial_2-\Delta^{-1}kx^1\dd x^3, 
       \Delta\partial_3+\Delta^{-1}kx^1\dd x^2; 
   dx^1, \Delta^{-1}dx^2, \Delta^{-1}dx^3) \ .   
\end{equation}

However $O_-$ and hence $K$ are clearly not single-valued, since
$x^1$ is periodic. Thus although locally we can gauge the $Q$
in~\eqref{eq:q-alg} away (and replace it with $f$ and $H$) we cannot
do this globally.  

This background is the simplest example of non-geometrical
compactification, where the $T^2$ fibres, labeled by $x^2$ and $x^3$,
are patched by a T-duality as one moves around the base $S^1$, labeled
by $x^1$. As such, it is not a manifold since T-duality does not map
points to points on the fibres. It is therefore hard to define in what
sense the basis $\ttcE_A$ is global. Nonetheless, the base $S^1$ is
still a conventional manifold, and we can simply imagine restricting
everything to this $S^1$ (or equivalently, ignoring the fact that the
fibres are compact). The $\ttcE_A$ are then a global basis for
the restricted generalized tangent space over $S^1$. We also have the
restrictions of $TM$ and $T^*M$. The statement that $Q$ cannot be
gauged away then has a well defined meaning in terms of the
restrictions, even if we cannot define the full compactification as a
manifold. 

We have seen that each of the three backgrounds, related by T-duality,
are parallelizable in the sense that one can introduce a globally
defined basis $\mathcal{E}_A$. The three
algebras~\eqref{eq:H-alg},~\eqref{ex-two} and~\eqref{eq:q-alg} are
actually equivalent: the only difference is the split of the basis
into $\mathcal{E}_a$ and $\mathcal{E}^a$, which is related to how $TM$
and $T^*M$ embed in $E$. The non-geometry of~\eqref{eq:q-alg} was
encoded in the fact that the $Q$-charge could not be gauged
away. Equivalently $E$, or rather its restriction to the base $S^1$,
could not be viewed as an extension of $TM$ by $T^*M$. Put another
way, its structure group was not in the $\Ggeom$ subgroup of
$O(3,3)$.

\subsubsection*{The Lunin-Maldacena solution}

The Lunin-Maldacena solution corresponds to a deformation of $AdS_5 \times S^5$
that was originally obtained  by applying a T-duality, a rotation and a
further T-duality on a $T^2$ inside $S^5$ \cite{LM}. 
$AdS_5 \times S^5$ can be written as a warped product of 4-dimensional Minkowski 
and the 6-dimensional flat metric. 
Defining the three complex coordinated on $\mathbb{R}^6$ as $z^i = \mu_i e^{ i \phi}$ \footnote{The coordinates $\mu_i$ are defined in terms of angles:
\beq
\mu_1 = \cos \alpha, \qquad \mu_2 = \sin \alpha \cos \theta, \qquad  
\mu_3 = \sin \alpha \sin \theta,
\eeq}, 
the  6-dimensional metric can be written as a (trivial) $T^3$ fibration
\beq
\label{metricR6}
{\rm d} s^2 = \sum_{i = 1}^3 ({\rm d} \mu_i)^2 + \mu_i^2 ({\rm d} \phi^i)^2 \, .
\eeq

As shown in \cite{MPZ, HT}, in this notation,  the chain
of transformations leading to the LM background is equivalent to a $\beta$-deformation.
In particular one can act on the generalized vielbein with the 
$\beta$-transform \eqref{eq:betatransform}, where
\beq
\label{betaLM}
\beta = \gamma \begin{pmatrix}
0 & 1 & -1 \\
-1 & 0 & 1 \\
1 & -1 & 0
\end{pmatrix} \, ,
\eeq
where $\gamma$ is the deformation parameter. 
Explicitely
\beq
\cE'= O \cE = \begin{pmatrix}
\mathbb{1} & & &\\
& \mathbb{1} & & \beta \\
& & \mathbb{1} &\\
& & & \mathbb{1}
\end{pmatrix} \, 
\begin{pmatrix} e_B & & & \\
                      & e_F & & \\
                      & & \hat{e}_B & \\
                      & & & \hat{e}_F \end{pmatrix}
 = 
\begin{pmatrix} e_B & & & \\
                      & e_F & & e_F \beta \\
                      & & \hat{e}_B & \\
                      & & & \hat{e}_F
\end{pmatrix} \, ,
\eeq
where $e_B = e^{a'} = \delta^{a'}{}_i {\rm d} \mu_i$ and
$e_F= e^a = \delta^a{}_i \mu_i  {\rm d} \phi^i$ are the vielbeins of the flat 
metric \eqref{metricR6} and 
\beq
e_F \beta = \gamma \begin{pmatrix}
0 & \mu_1 & -\mu_1 \\
-\mu_2 & 0 & \mu_2 \\
\mu_3 & - \mu_3 & 0
\end{pmatrix} \, .
\eeq

From the generalized metric $\mathcal{H}$ it is easy to see that the new metric and $B$-field
are indeed those of the LM solution
\bea
\label{LMsol}
&& {\rm d}s^2 =\sum_{i=1}^3 ({\rm d} \mu_i + G {\rm d}\phi^i)^2 + \gamma^2 G 
(\mu_1 \mu_2 \mu_3)^2 (\sum_{i=1}^3 {\rm d}\phi^i)^2  \, ,\\
&& B = \gamma G [ (\mu_1 \mu_2 )^2 {\rm d}\phi^1 \wedge {\rm d}\phi^2 +
(\mu_2 \mu_3 )^2 
 {\rm d}\phi^2 \wedge  {\rm d}\phi^3 
+ (\mu_3 \mu_1 )^2 {\rm d}\phi^3 \wedge  {\rm d}\phi^1] \, ,
\eea
with $G = [1 +\gamma^2 ((\mu_1 \mu_2 )^2 + (\mu_1 \mu_3 )^2  + (\mu_2 \mu_3 )^2)]^{-1}$.

As in the previous example we can find an $O(d) \times O(d)$ transformation 
bringing the generalized vielbein to the triangular form \eqref{eq:genvielbeinm}. In this
case \eqref{eq:betacomp} takes the form
\beq
O_+ = \mathbb{I} \ \,  \qquad \qquad O_- = \begin{pmatrix} 1 & \\ & G O^{\rm F}_-
\end{pmatrix}
\eeq
with
\beq
 O^{\rm F}_-  = \begin{pmatrix} 
1 - \gamma^2 (\mu_1^2 \mu_2^2 - \mu_2^2 \mu_3^2 + \mu_1^2 \mu_3^2) 
& 2 \gamma \mu_1 \mu_2 ( 1 + \gamma \mu_3) & 2 \gamma \mu_1 \mu_3 ( - 1 + \gamma \mu_2) \\
 2 \gamma \mu_1 \mu_2 ( - 1 + \gamma \mu_3) & 1 - \gamma^2 (\mu_1^2 \mu_2^2  + \mu_2^2 \mu_3^2 
- \mu_1^2 \mu_3^2) & 2 \gamma \mu_2 \mu_3 ( 1 + \gamma \mu_1) \\
2 \gamma \mu_1 \mu_3 ( 1 + \gamma \mu_2) & 2 \gamma \mu_2 \mu_3 ( - 1 + \gamma \mu_1) &
1 - \gamma^2 ( \mu_2^2 \mu_3^2 + \mu_1^2 \mu_3^2 - \mu_1^2 \mu_2^2 ) 
\end{pmatrix}
\eeq

Differently from the previous example, the transformation $O_-$ does
not contain any non-single valued function of the base. This is also related to the
fact that since we have a simply connected base it is not possible to choose a $B$-field
with two legs along the fibre to be not single valued.

\subsection{Generalized parallelizable backgrounds}
\label{sec:gen-para}

The simplest way around the gauge-dependence of the charges $F$ is to
assume that there is some preferred frame $\mathcal{E}_A$, and to
define $F$ as the values in this frame. In the previous examples, such
a class of frames was defined by the $\mathbb{T}^d$ fibration
structure. In particular, for those based on the three-torus with
$H$-flux, there was actually a fixed globally defined frame. This is
an example of a ``generalized parallelizable'' background. In this
section, we would like briefly to address some of the constraints on
the generic form of the local geometry of such backgrounds, and
in particular ask what charges $F$ can appear. We will also see how
T-duality acts on such backgrounds. 

Recall that in conventional geometry on a parallelizable manifold
there exists a globally defined frame $e^a$ implying the tangent
bundle $TM$ is trivial. In addition one can further assume that the
manifold admits a metric of the form $g=g_{ab}e^a\otimes e^b$ with
$g_{ab}$ constant. (In the mathematics literature this is known as
``consistent absolute parallelism''~\cite{JFF,Wolf}.) Except for the
special case of $S^7$, the manifold is then a Lie group and the
functions $f^a{}_{bc}$ are the structure constants. In complete
analogy one can define a  ``generalized parallelizable
compactification'' where there is now a globally defined frame 
$\mathcal{E}_A$ of $E$.  We will also  assume that the $O(d,d)$ metric
takes the form~\eqref{eq:Om-basis}, which we can also write as  
\begin{equation}
   \eta = \eta^{AB} \cE_A \otimes \cE_B \ ,
\end{equation}
but drop the requirement that $\mathcal{H}$ takes a
particular form. Thus the $\cE^A$ are defined up to global $O(d,d)$
transformations. Up to such rotations, there is then a unique set of
charges defined by  
\begin{equation}
\label{eq:GPC}
   [ \cE_A , \cE_B ] = F^C{}_{AB} \cE_C \, ,
\end{equation}
which are taken to be constant. Again, the notion of ``globally
defined'' becomes unclear when we talk about non-geometrical
backgrounds. As it stands we will only assume such a local geometry
and corresponding charges. The question of how these might complete
into geometrical or non-geometrical backgrounds is not discussed. Note
that such backgrounds are somewhat analogous to the general twisted
double torus backgrounds discussed for instance
in~\cite{HRE,dallprezas}. The difference is that there the algebra is
realized in terms of the Lie bracket of vector fields on a ``doubled'' 
$2d$-dimensional space. Here we are considering a more restricted
example: we use the Courant bracket on generalized vectors on what is
locally a conventional $d$-dimensional space. 

The three-torus examples above are each generalized parallelizable
manifolds. The algebras of the $\cE_A$ are actually isomorphic in each
case. It is the split of $E$ into $TM$ and $T^*M$ (and hence of
$\cE_A$ into $(\cE_a,\cE^a)$) that gave the different interpretations
of the structure constants as corresponding to $H$, $f$ or $Q$ charge. 

Let us see what conditions the existence of the algebra~\eqref{eq:GPC}
realized by the Courant bracket places on the local geometry of the
background. The first conditions follow from the fact that we
can define the $O(d,d)$ metric ${\cal H}$ as in~\eqref{eq:bases}. From
Proposition~3.16 of~\cite{Gualtieri} we see that, since the
$F^A{}_{BC}$ are constant, the Courant bracket~\eqref{eq:GPC} on
$\cE_A$ satisfies the Jacobi identity and hence defines a Lie
algebra $\mathfrak{h}$. Given Proposition~3.18 of~\cite{Gualtieri}, we
also have  
\begin{equation}
   \eta_{CD} F^D{}_{AB} + \eta_{BD}F^D{}_{AC} = 0 \ .
\end{equation}
This implies that the adjoint representation of the
algebra~\eqref{eq:gen-alg}, where the generators are given by
$(T_A)_B{}^C=F^C{}_{AB}$, acts as a sub-algebra
$\mathfrak{h}\subset\mathfrak{o}(d,d)$.

Next recall that under the projection $\pi:E\to TM$ the Courant bracket
reduces to the Lie bracket  
\begin{equation}
   \pi([X,Y]) = [\pi(X),\pi(Y)]_\text{Lie} \ . 
\end{equation}
Writing $v_A=\pi(\cE_A)$ this simply states that
$[v_A,v_B]_\text{Lie}=f^C{}_{AB}v_C$. Thus there is a realization of
the algebra $\mathfrak{h}$ in terms of $2d$ vector fields on $M$, though of
course this may be somewhat degenerate since some $v_A$ may vanish
identically. Since the $\cE_A$ are a basis for $E$, the $v_A$ must
form a basis for $TM$, that is, there must be at least $d$
non-vanishing $v_A$ at each point $p$ of $M$.  Exponentiating the Lie
algebra action into diffeomorphisms we see that $M$ is locally a
homogeneous space, with an action of a group $H$ with Lie algebra
$\mathfrak{h}$. Let us fix some point $p\in M$. If we identify
$X=X^A\cE_A$ with constant $X^A$ as elements of the Lie algebra
$\mathfrak{h}$ we define the set of vectors $X$ with vanishing
$\pi(X)$ at a given point $p\in M$  
\begin{equation}
   \mathfrak{k}_p = \left\{ X \in \mathfrak{h} : \pi(X)|_p=0 \right\}
   \ .
\end{equation}
This must be a $d$-dimensional subset of $\mathfrak{h}$. Since the Lie
bracket of two vector fields that vanish at $p\in M$ must itself
vanish at $p\in M$, we see that $\mathfrak{k}_p$ must form a closed
subalgebra. Hence we see that locally $M$ must be a coset space. We
can write  
\begin{equation}
   M = H/K \ , \qquad K \subset H \subset O(d,d) \ , 
\end{equation}
where $H$ is a $2d$-dimensional group with Lie algebra $\mathfrak{h}$
given by~\eqref{eq:GPC} and $K$ is a $d$-dimensional subgroup with Lie 
algebra isomorphic to $\mathfrak{k}_p$. For a parallelizable manifold,
$M$ is (almost always) locally a group manifold. Thus we see, as one
might expect, generalized parallelizable compactification appear to be
more general. 

Let us now turn to the fluxes $F$. At the point $p$, generalized
vectors $X\in\mathfrak{k}_p$ lie solely in $T^*_pM$. Hence we can
locally identify $\cE^a$ as a basis for $\mathfrak{k}_p$ and, using
the metric $\eta$, decompose
$\mathfrak{h}=\mathfrak{k}_p\oplus\mathfrak{m}_p$, with $\cE_a$ a
basis for $\mathfrak{m}_p$. Hence, for any generalized parallelizable
compactification, since $\mathfrak{k}_p$ is a closed subalgebra, we
see that one cannot arrange all fluxes to be non-zero. In particular, 
one can always use a global $O(d,d)$ rotation to align the basis
$\cE_A$ such that $\cE^a$ span $\mathfrak{k}_p$ and $\cE_A$ span
$\mathfrak{m}_p$ and 
\begin{equation}
   R^{abc} = 0 \ . 
\end{equation}
This is in agreement with the $T^3$ with flux examples discussed
above. 

In this discussion we have only considered some of the conditions on
the generalized parallelizable background that follow from the Courant
bracket structure. One would expect additional conditions, such as
compatibility with a generalized metric of the form
$\mathcal{H}=\mathcal{H}^{AB}\cE_A\otimes\cE_B$, and probably a
curvature condition as in~\cite{dallprezas}. It would also be interesting to
find specific examples where $M$ is indeed locally a coset rather than
a group manifold as in the $T^3$ with $H$-flux examples. 

Let us end this section by discussing the generic action of T-duality
on generalized parallelizable backgrounds. Suppose we have a
generalized Killing vector $V$ which preserves the parallelizable
structure, that is 
\begin{equation}
\label{eq:Vpara}
   \Lgen_V \cE_A = 0 \qquad \forall A\ . 
\end{equation}
We can always normalize $V$ such that $\eta(V,V)=1$ and define the
T-duality operator $T_V$ as in~\eqref{eq:TVdef}. Using the general
relations~\cite{Gualtieri}
\begin{equation}
\begin{gathered}
   \Lgen_X Y = [X,Y] + \dd \eta(X,Y) \ , \\
   i_{\pi(X)}\dd\eta(Y,Z) = \eta(\Lgen_XY,Z) + \eta(Y,\Lgen_XZ) \ , \\
   [X,fY] = f[X,Y] + (i_{\pi(X)}\dd f)Y - \eta(X,Y)\dd f \ ,
\end{gathered}
\end{equation}
and the fact that $\eta(\cE_A,\cE_B)=\eta_{AB}$ and $\eta(V,V)=1$ are
constant, it is relatively straightforward to show that 
\begin{equation}
\begin{aligned}
   {}[T_V\cE_A,T_V\cE_B] &= [ \cE_A, \cE_B ] 
      - 2\left(i_{\pi(\cE_A)}\dd\eta(V,\cE_B)
           - i_{\pi(\cE_B)}\dd\eta(V,\cE_A) \right) V \\
      &= T_V[\cE_A,\cE_B] \ . 
\end{aligned}
\end{equation}
Thus we see that $T_V$ is an automorphism of the generalized
parallelizable algebra. A particular example is the fact that the
three algebras arising from T-duality of the $T^3$ with $H$-flux
are all isomorphic. They are of course distinguished by the
way one identifies vectors and forms in $E$.





\subsubsection*{Acknowledgments}

We thank Dima Belov,  Gianguido D'Allagata, Nick Halmagyi and Andrei Micu for useful discussions. This work is supported in part by RTN contracts  MRTN-CT-2004-005104 and  MRTN-CT-2004-512194 and by ANR grants BLAN06-3-137168 (MG and RM) and BLAN05-0079-01 (MP).


\appendix





\end{document}